\documentclass[9pt,twocolumn,twoside]{osajnl}

\journal{ao} 
\usepackage{graphicx}
\usepackage{textcomp}
\usepackage{subfigure}
\usepackage{amssymb}
\usepackage{amsfonts}
\usepackage{csquotes}
\usepackage{color}
\usepackage{array}
\usepackage{braket}
\usepackage{booktabs}
\setboolean{shortarticle}{true}

\title{Joint Power and Gain Allocation in MDM-WDM Optical Communication Networks Based on Enhanced Gaussian Noise Model}
 
\author[1]{M. A. Amirabadi}
\author[1,*]{M. H. Kahaei}
\author[1]{S. A. Nezamalhosseini}

\affil[1]{School of Electrical Engineering, Iran University of Science and Technology (IUST), Tehran, 1684613114 Iran}

\affil[*]{Corresponding author: kahaei@iust.ac.ir}

\begin{abstract}
Achieving reliable communication over different channels and modes is one of the main goals of Mode Division Multiplexing-Wavelength Division Multiplexing (MDM-WDM) communication networks. The reliability can be described by minimum Signal to Noise Ratio (SNR) margin which dependents on launched power, Multimode-Erbium Doped Fiber Amplification (MM-EDFA) gain, and MMF nonlinearity.
In this paper, an analytical model for MMF nonlinearity is derived based on Enhanced Gaussian Noise (EGN) model formulation by considering carrier phase estimation and the first four dispersion terms. The proposed EGN model is verified through the split step Fourier method simulation.
Considering a multi-node linear network, the joint optimized power and gain allocation based on minimum SNR margin maximization is formulated.
The practical constraints including MM-EDFA saturation power and maximum gain are considered and the problem is solved by using convex optimization.
Three scenarios are assumed including best equal power, optimized power, and joint optimized power and gain.
In the first scenario, equal powers are considered for different channels and modes with equal MM-EDFA gain in all spans. It is worth mentioning that the MM-EDFA gain is equal to span loss.
In the second scenario, different powers are allocated to different channels and modes with equal MM-EDFA gain in all spans.
In the third case, allocated powers to each channel and mode are optimized. Moreover, the MM-EDFA gain for each span is optimized separately.
In MDM-single channel systems, simulation results demonstrate that joint power and gain optimization leads to {\small $0.61~dB$} and {\small $0.62~dB$} minimum SNR margin improvement compared to optimal power and best equal power allocation, respectively. In SMF-WDM systems considering the joint optimal power and gain allocation, the minimum SNR margin improvements is about {\small $1.72~dB$} and {\small $1.71~dB$} in comparison with optimal power and best equal power allocation, respectively.
\end{abstract}

\setboolean{displaycopyright}{true}

\begin{document}

\maketitle

\section{Introduction}
Mode Division Multiplexing (MDM) over Multimode Fibers (MMF) or multicore fibers has emerged as a possible solution for overcoming the data-rate crunch in optical communication networks. Theoretically, deploying an MMF propagating {\small $D$} spatial modes would increase capacity {\small $D$} times \cite{1}. The combination of MDM with Wavelength Division Multiplexing (WDM) and polarization division multiplexing schemes further increases the data rate.

The MDM-WDM systems suffer from both MMF linear and nonlinear effects. The MMF linear effects include attenuation, chromatic and modal dispersion, and linear coupling. The MMF nonlinear impairments include Kerr-effect-based nonlinearity and nonlinear coupling \cite{1}. The chromatic and modal dispersion, as well as the linear coupling, can be mitigated by Multi-Input Multi-Output (MIMO) Digital Signal Processing (DSP) techniques \cite{2}. However, the MMF nonlinear effects are the main limitations towards the practical implementation of MDM-WDM systems.

During the last decade, several theoretical works have considered MMF nonlinearity analysis which focused mainly on numerical simulations \cite{3} and analytical predictions \cite{9} combined with experimental verifications \cite{11}.
The first step towards analyzing the MMF nonlinearity is solving the Manakov equation \cite{3}.
The Split Step Fourier Method (SSFM) solves the Manakov equation through many successive numerical simulation steps. The SSFM has high computational complexity. The perturbation-based methods are the mostly used analytical models that approximately solve the Manakov equation \cite{13},\cite{16}, and result in analytical formulations for predicting MMF nonlinearity. The Gaussian Noise (GN) model is the most practical perturbation-based model which describes the nonlinear effects by an additive Gaussian noise source \cite{17}, \cite{18}.
The GN model is accurate only for Gaussian shaped constellations in long-range multi-span links.

In the first part of the paper, the Enhanced GN (EGN) model is derived by considering Carrier Phase Estimation (CPE) and the first four dispersion terms in order to improve the model accuracy. Moreover, the accuracy of the proposed model is validated by SSFM.
The proposed EGN formulation provides a clear relationship between MMF nonlinearity with launched power of different channels and modes as well as the MM-Erbium Doped Fiber Amplifier (MM-EDFA) gain. Furthermore, the MM-EDFA produces the ASE noise which can be modeled by an additive white Gaussian noise. Therefore, the joint power and gain allocation problem can be composed by considering the ASE noise and nonlinear noise formulations.

Achieving reliable communication is one of the main goals of MDM-WDM systems which can be described by the minimum SNR margin. Therefore, the joint optimized power and gain allocation considering minimum SNR margin maximization is presented in the second part of the paper.
Some practical constraints including MM-EDFA saturation power and MM-EDFA maximum gain are considered. The problem is solved considering a multi-node linear network by using convex optimization approaches. The rest of this paper is organized as follows. Section II presents the system and signal model, Section III provides the EGN model formulation, Section IV presents the problem statement, Section V brings the simulation results, and Section VI concludes the paper.

\begin{figure}[tp!]
    \centering
\includegraphics[width=0.9\linewidth, height=4.5cm]{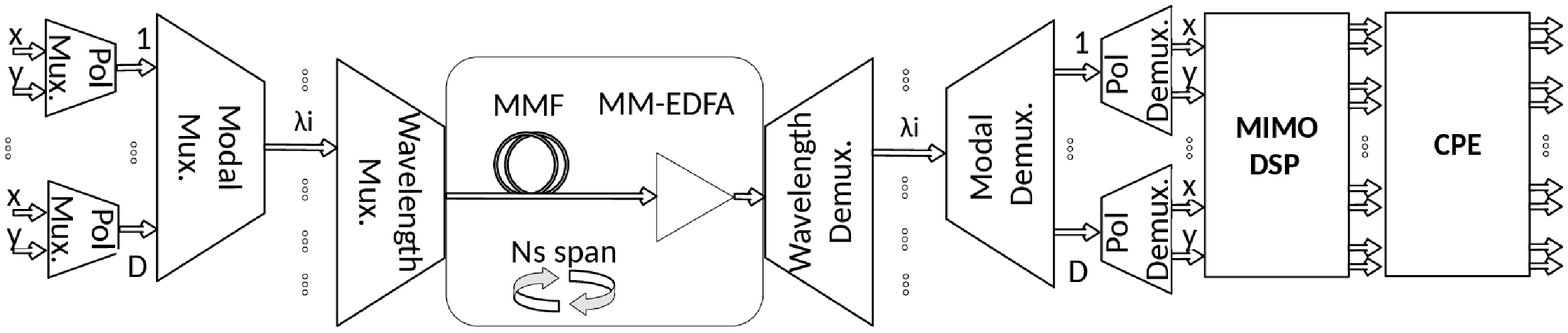}
\caption{Schematic diagram of the MDM-WDM system.}
\label{fig:1}
\end{figure}

\section{System and signal model}
\subsection{System model}
Fig. 1 shows the considered MDM-WDM system in which the input data is a multiplexing of {\small $N_{ch}$} channels, {\small $D$} spatial modes, and {\small $2$} polarization modes. This link has {\small $N_s$} spans with length {\small $L_s$}, combined by an MM-EDFA at the end of each span for compensating the optical fiber loss.
In order to minimize the MMF nonlinear effect, both modal and chromatic dispersion are not compensated \cite{17}.
The Kerr nonlinearity produced by inter/intra channel and mode interactions is considered.
It is assumed that the MMF imperfections causing linear coupling among spatial modes be so weak, therefore, the linear coupling effect is negligible \cite{3}.
Moreover, a MIMO DSP at the receiver is considered for compensating MMF linear effects. Following, a CPE is used to recover the average carrier phase rotation \cite{16} due to the MMF nonlinear effect.

\subsection{Signal model}
The following Ket notation represents the time domain of the optical launched signal into the MMF link as
\small{\begin{equation}\label{eq:1}\begin{aligned}
\ket{A(0,t)}=\sum_{i_1=-\infty}^{\infty} \sum_{i_2=1}^{N_{ch}} \sum_{i_3=1}^{2D} \zeta_{\mathbf{i}}W_{Tx}^{i_2,i_3}(t-i_1T_{i_2}) e^{j2\pi f_{i_2}t} \ket{i_3},
\end{aligned}\end{equation}}	
where {\small $\zeta_{\mathbf{i}}$} is the digital symbol (for example, QPSK) at time index {\small $i_1$}, WDM channel index {\small $i_2$}, polarization-spatial mode index {\small $i_3;i_3=1,…,2D$}, and {\small $\mathbf{i}=[i_1,i_2,i_3]$}. Moreover, {\small $W_{Tx}^{i_2,i_3}(t-i_1 T_{i_2})$} is the transmitted pulse at time index {\small $i_1$}. Moreover, {\small $i_2$} and {\small $i_3$} represent WDM channel and polarization-spatial mode index, respectively.
{\small $T_{i_2}$} shows the symbol duration, and {\small $f_{i_2}$} is the carrier frequency. {\small $\ket{i_3} $} represents a one-hot vector wherein the {\small$i_3 $}th element is one and the other elements are zero.
Actually, {\small $i_3$} is used to denote the polarization-spatial mode of the propagated signal. In other words, {\small $i_3\triangleq p^{'},p$}, where {\small $p^{'};p^{'}=x,y$} is the polarization mode index and {\small $p;p=1,…,D$} represents the spatial mode index. It is obvious that {\small $i_3$} takes values between {\small $1$} and {\small $2D$}, since each spatial mode is a multiplexing of {\small $2 $} polarization modes.
Therefore, the time domain of the  optical launched signal into the MMF can be expressed as
\small{\begin{equation}\label{eq:2}\begin{aligned}
{}&\ket{A(0,t)} = \sum_{i_3=1}^{2D} A_{i_3}(0,t) \ket{i_3}\triangleq \begin{bmatrix}
A_{1}(0,t)\\
A_{2} (0,t)\\
…\\
A_{2D-1}(0,t) \\
A_{2D}(0,t)
\end{bmatrix}
\triangleq
\begin{bmatrix}
A_{x,1}(0,t)\\
A_{y,1} (0,t)\\
…\\
A_{x,D}(0,t) \\
A_{y,D}(0,t)
\end{bmatrix},
\end{aligned}\end{equation}}
and
\small{\begin{equation}\label{eq:3}\begin{aligned}
{}&\bra{A(0,t) } \triangleq \begin{bmatrix}
A_{1}^* (0,t) & A_{2}^* (0,t) & … & A_{2D-1}^* (0,t) & A_{2D}^* (0,t)
\end{bmatrix} \\
&\triangleq \begin{bmatrix}
A_{x,1}^* (0,t) & A_{y,1}^* (0,t) & … & A_{x,D}^* (0,t) & A_{y,D}^* (0,t)
\end{bmatrix},
\end{aligned}\end{equation}}
where {\small $ A_{i_3}(0,t) \triangleq A_{p^{'},p}(0,t)$} represents the time domain of the propagated signal in {\small $ i_3$}th polarization-spatial mode (i.e., {\small $p^{'}$}th polarization mode and {\small $p$}th spatial mode) and can be expressed as

\small{\begin{equation}\label{eq:4}\begin{aligned}
\ket{A_{i_3}(0,t)}=\sum_{i_1=-\infty}^{\infty} \sum_{i_2=1}^{N_{ch}} \zeta_{\mathbf{i}}W_{Tx}^{i_2,i_3}(t-i_1T_{i_2}) e^{j2\pi f_{i_2}t} \ket{i_3},
\end{aligned}\end{equation}}

Considering (\ref{eq:1}), the propagated signal in the frequency domain can be expressed by
\small{\begin{equation}\label{eq:5}\begin{aligned}
\ket{\tilde{A}(0,f)}=\sum_{\mathbf{i}} \zeta_{\mathbf{i}} \ket{\tilde{W}_{Tx}^{\mathbf{i}}(f)},
\end{aligned}\end{equation}}
where
\small{\begin{equation}\label{eq:6}\begin{aligned}
{}&\ket{\tilde{W}_{Tx}^{\mathbf{i}}(f)} \triangleq \tilde{W}_{Tx}^{\mathbf{i}}(f) \ket{i_3} =\tilde{W}_{Tx}^{i_2,i_3}(f-f_{i_2})e^{-j2\pi(f-f_{i_2})i_1T_{i_2}}\ket{i_3}.
\end{aligned}\end{equation}}

\newcounter{MYtempeqncnt}
\begin{figure*}[!t]
\normalsize
\setcounter{MYtempeqncnt}{\value{equation}}
\setcounter{equation}{16}
\small{\begin{equation}\label{eq:17}\begin{aligned}
n_\mathbf{i}(f)=-j\sum_{\mathbf{{\mathbf{k,m,n}}}}\zeta_\mathbf{k}^*\zeta_\mathbf{m}\zeta_\mathbf{n} \iiint_{-\infty}^{\infty} \ket{\eta(f,f_1,f_2)} \braket{\tilde{W}_{Tx}^{\mathbf{k}}(f+f_1+f_2)|\tilde{W}_{Tx}^{\mathbf{m}}(f+f_2)} \braket{g^{\mathbf{i}}(f)|\tilde{W}_{Tx}^{\mathbf{n}}(f+f_1)} R_\mathbf{i} df_1  df_2 df,
\end{aligned}\end{equation}}
\setcounter{equation}{\value{MYtempeqncnt}}
\hrulefill
\vspace*{4pt}
\end{figure*}
\newcounter{MYtempeqncnt1}
\begin{figure*}[!t]
\normalsize
\setcounter{MYtempeqncnt}{\value{equation}}
\setcounter{equation}{17}

\small{\begin{equation}\label{eq:18}\begin{aligned}
n_\mathbf{i}(f)=-j\sum_{\mathbf{k,m,n}}\zeta_\mathbf{k}^*\zeta_\mathbf{m}\zeta_\mathbf{n}\iiint_{-\infty}^{\infty} \ket{\eta(f,f_1,f_2)}
\tilde{W}_{Tx}^{{\mathbf{k}}^*}(f+f_1+f_2) \tilde{W}_{Tx}^{\mathbf{m}}(f+f_2) g^{{\mathbf{i}}^*}(f) \tilde{W}_{Tx}^{\mathbf{n}}(f+f_1) \braket{k_3|m_3} \braket{i_3|n_3} R_\mathbf{i} df_1 df_2 df,
\end{aligned}\end{equation}}
\setcounter{equation}{\value{MYtempeqncnt}}
\hrulefill
\vspace*{4pt}
\end{figure*}
\newcounter{MYtempeqncnt2}
\begin{figure*}[!t]
\normalsize
\setcounter{MYtempeqncnt}{\value{equation}}
\setcounter{equation}{18}

\small{\begin{equation}\label{eq:19}\begin{aligned}
{}&E[n_{\mathbf{i}}(f) n_{\mathbf{i}}^*(f)]=\sum_{\mathbf{kmnljo}} E[\zeta_{\mathbf{k}}^*\zeta_{\mathbf{m}}\zeta_{\mathbf{n}}\zeta_{\mathbf{l}}\zeta_{\mathbf{j}}^*\zeta_{\mathbf{o}}^*] \idotsint_{\infty}^{\infty}\braket{\eta(f,f_1,f_2) |\eta(v,v_1,v_2)} \tilde{W}_{Tx}^{\mathbf{k}*}(f+f_1+f_2)\tilde{W}_{Tx}^{\mathbf{m}}(f+f_2)\\
&g^{\mathbf{i}*}(f)\tilde{W}_{Tx}^{\mathbf{n}}(f+f_1) \tilde{W}_{Tx}^{\mathbf{j}*}(v+v_2)\tilde{W}_{Tx}^{\mathbf{l}}(v+v_1+v_2) \tilde{W}_{Tx}^{\mathbf{o}*}(v+v_1)g^{\mathbf{i}}(v) \braket{k_3|m_3} \braket{i_3|n_3}\braket{j_3|l_3} \braket{o_3|i_3} R_\mathbf{i} df_1  df_2 df dv_1  dv_2 dv.
\end{aligned}\end{equation}}
\setcounter{equation}{\value{MYtempeqncnt}}
\hrulefill
\vspace*{4pt}
\end{figure*}
\newcounter{MYtempeqncnt19}
\begin{figure*}[!t]
\normalsize
\setcounter{MYtempeqncnt}{\value{equation}}
\setcounter{equation}{24}

\small{\begin{equation}\label{eq:25}\begin{aligned}
{}&\sigma_{EGN,i_2,p}^2 = \sum_{s=1}^{N_s}\prod_{n=1}^{s-1}(G_n~L_n)^3\prod_{n=s}^{N_s}(G_s~L_s) \sum_{q=1}^{D}\bigg[3/4\sum_{k_2,m_2,n_2}\kappa_1^{(k_2)}\kappa_1^{(m_2)}\kappa_1^{(n_2)} P_{k_2,q} P_{m_2,q} P_{n_2,p}X_{l_2,p}^{a}(k_2,m_2,n_2,q)+\\
&1/4\sum_{k_2,n_2}\kappa_2^{(k_2)} \kappa_1^{(n_2)}  (P_{k_2,q}^2 P_{n_2,p}5X_{l_2,p}^{b}(k_2,k_2,n_2,q)+P_{k_2,p} P_{k_2,q}P_{n_2,q}X_{l_2,p}^{c}(k_2,n_2,k_2,q))+\\
& 1/4\sum_{n_2}\kappa_3^{(n_2)}  P_{n_2,q}^2P_{n_2,p} X_{l_2,p}^{d}(n_2,n_2,n_2,q)\bigg]
\end{aligned}\end{equation}}
\setcounter{equation}{\value{MYtempeqncnt}}
\hrulefill
\vspace*{4pt}
\end{figure*}

\section{EGN model formulation}
The Manakov equation for the considered MDM-WDM link can be expressed as \cite{16}
\small{\begin{equation}\label{eq:7}\begin{aligned}
{} & \frac{\partial \ket{A_{i_3}(z,t)}}{\partial z}  = \mathcal{L}+\mathcal{N}
\end{aligned}\end{equation}}
where the linear and nonlinear effects of MMF can be shown as \cite{3}
\small{\begin{equation}\label{eq:8}\begin{aligned}
{}&\mathcal{L}=-\frac{\alpha_{i_3}}{2}\ket{A_{i_3}(z,t)} +j {\beta_0}_{i_3} \ket{A_{i_3}(z,t)} - {\beta_1}_{i_3}\frac{\partial \ket{A_{i_3}(z,t)}}{\partial t}\\
& -j\frac{{\beta_2}_{i_3}}{2}\frac{\partial^2 \ket{A_{i_3}(z,t)}}{\partial t^2} -\frac{{\beta_3}_{i_3}}{6}\frac{\partial^3 \ket{A_{i_3}(z,t)}}{\partial t^3}
\end{aligned}\end{equation}}
\small{\begin{equation}\label{eq:9}\begin{aligned}
{}&\mathcal{N} = j \sum_{k_3,m_3,n_3=1}^{2D} \tilde{\gamma}_{i_3k_3m_3n_3} \bigg( \braket{A_{n_3}(z,t)|A_{m_3}(z,t)} \ket{A_{k_3}^{*} (z,t)}\\
&+\braket{A_{k_3}^{*}(z,t)|A_{m_3} (z,t)} \ket{A_{n_3} (z,t)}\bigg) ,
\end{aligned}\end{equation}}
{\small $\alpha_{i_3}$} is the attenuation of {\small $i_3$}th polarization-spatial mode. {\small $\tilde{\gamma}_{i_3i_3i_3i_3}=\frac{8}{9}\gamma f_{i_3i_3i_3i_3}$}, and {\small $\tilde{\gamma}_{i_3k_3m_3n_3}=\frac{4}{3}\gamma f_{i_3k_3m_3n_3}$} where {\small $\gamma$} is the Kerr nonlinearity coefficient, with {\small $f_{i_3k_3m_3n_3}=\frac{A_{eff}}{\sqrt{I_{i_3}I_{k_3}I_{m_3}I_{n_3}}} \iint{F_{i_3}(x,y) F_{k_3}(x,y) F_{m_3}(x,y) F_{n_3}(x,y) dx dy}$} is the nonlinear coupling coefficient between modes {\small $i_3$}, {\small $k_3$}, {\small $m_3$} and {\small $n_3$} where {\small $F_{i_3}(x,y)$} is the spatial profile of {\small $i_3$}th mode, {\small $I_{i_3}=\iint{F_{i_3}^2(x,y)dxdy}$}, {\small $A_{eff}$} is the effective area of the fundamental mode \cite{17}.

The first-order perturbation approximation of Manakov equation solution expresses the received signal as \cite{16}
\small{\begin{equation}\label{eq:10}\begin{aligned}
\ket{A(z,t)}\simeq e^{\mathcal{L}z}\ket{A(0,t)}+\int_0^z e^{\mathcal{L} (z-\xi)}\mathcal{N} \bigg(e^{\mathcal{L}\xi}\ket{A(0,t)} \bigg) d\xi,
 \end{aligned}\end{equation}}

After compensating the linear effects of MMF using the MIMO DSP, the received signal can be expressed as
\small{\begin{equation}\label{eq:11}\begin{aligned}
\ket{A_R}\simeq \ket{A(0,t)}+\int_0^z e^{-\mathcal{L}\xi}\mathcal{N} (e^{\mathcal{L}\xi}\ket{A(0,t)})d\xi,
 \end{aligned}\end{equation}}

The {\small $\mathcal{L}$} can be described in frequency domain by its Fourier transform {\small $\mathcal{F}(e^{\mathcal{L}z})= \ket{e^{\nu(z,f)}}$} where
\small{\begin{equation}\label{eq:12}\begin{aligned}
\nu_{i_3}(z,f)=-\int_0^z(\alpha_{i_3}(\xi)+j\beta_{i_3}(\xi,f))d\xi,
\end{aligned}\end{equation}}
Moreover, {\small $\beta_{i_3}(z,f)$} can be calculated as
\small{\begin{equation}\label{eq:13}\begin{aligned}
\beta_{i_3}(z,f)=\beta_{0_{i_3}}+\beta_{1_{i_3}}(2\pi f)+\frac{\beta_{2_{i_3}}}{2}(2\pi f)^2+ \frac{\beta_{3_{i_3}}}{6}(2\pi f)^3.
\end{aligned}\end{equation}}

The second term in (\ref{eq:11}) represents the nonlinear noise which by considering (\ref{eq:12}) can be simplified  in the
frequency domain as
\small{\begin{equation}\label{eq:14}\begin{aligned}
{}&\ket{\tilde{n}(f)}=-j\iint_{-\infty}^{-\infty} \ket{\eta(f,f_1,f_2)}\braket{\tilde{A}(f+f_1+f_2)| \tilde{A}(f+f_2)}\\
& \ket{\tilde{A}(f+f_1)} df_1 df_2,
 \end{aligned}\end{equation}}
where
\small{\begin{equation}\label{eq:15}\begin{aligned}
{}&\eta_{i_3}(f,f_1,f_2)\triangleq \sum_{k_3,m_3,n_3} \tilde{\gamma}_{i_3,k_3,m_3,n_3} \\
&\int_0^z e^{\nu_{n_3}(\xi,f+f_1)+ \nu_{m_3} (\xi,f+f_2)+\nu_{k_3}^*(\xi,f+f_1+f_2)-\nu_{i_3} (\xi,f)}d\xi.
\end{aligned}\end{equation}}

The matched filtering deployment on the received signal can be described by
\small{\begin{equation}\label{eq:16}\begin{aligned}
\int_{-\infty}^{\infty}\braket{g^{\mathbf{i}}(f)|\tilde{A}(f)} R_\mathbf{i}df ,
\end{aligned}\end{equation}}
where {\small $\ket{g^{\mathbf{i}}(f)}= g^{\mathbf{i}}(f) \ket{i_3}$} is the spectral shape of transmitted pulse on {\small $i_2$}th channel and {\small $i_3$}th polarization-spatial mode which has been normalized such that {\small $\int_{-\infty}^{+\infty} g^{\mathbf{i}}(f) df=1$}, {\small $R_\mathbf{i}$} is the symbol rate of transmitted pulse on {\small $i_2$}th channel and {\small $i_3$}th polarization-spatial mode.
Accordingly, the nonlinear noise takes the form of (\ref{eq:17}) at the receiver which can be written as (\ref{eq:18}).

The variance of the nonlinear noise is derived in (\ref{eq:19}), which depends on six infinite integration/summations where the following Poisson summation helps dropping some interactions/summations:
\addtocounter{equation}{3}
\small{\begin{equation}\label{eq:20}\begin{aligned}
\sum_{k=-\infty}^{\infty}e^{j kf 2\pi T} = T \sum_{k=-\infty}^{\infty}\delta(f-k/T).
\end{aligned}\end{equation}}
By considering the {\small $sinc$} pulses with finite bandwidth interacting with only one Dirac's delta, summation over time index can be in (\ref{eq:19}). Furthermore, equating the arguments with equal atoms results in
\small{\begin{equation}\label{eq:21}\begin{aligned}
{}&f+f_1+f_2=v+v_1+v_2\\
&f+f_2=v+v_2\\
&f+f_1=v+v_1,
\end{aligned}\end{equation}}
and accordingly
\small{\begin{equation}\label{eq:22}\begin{aligned}
{}&f=v\\
& f_1=v_1\\
& f_2=v_2,
\end{aligned}\end{equation}}
which yields to simplify three integrals with {\small $v, v_1$}, and {\small $v_2$}.

\newcounter{MYtempeqncnt20}
\begin{figure*}[!t]
\normalsize
\setcounter{MYtempeqncnt}{\value{equation}}
\setcounter{equation}{27}

\small{\begin{equation}\label{eq:28}\begin{aligned}
{}& M_{i_2,p}= \Bigg(P_{i_2,p}\prod_{n=1}^{N_s}(G_n~L_n)/ \bigg( \prod_{n=1}^{N_s}(G_n~L_n) (F (G_{BA}-1) h \nu B_{i_2})+\sum_{s=1}^{N_s}[(F(G_s-1)h\nu B_{i_2})\prod_{n=s+1}^{N_s}(G_n~L_n)] +\sum_{s=1}^{N_s}\prod_{n=1}^{s-1}(G_n~L_n)^3\\
&\prod_{n=s}^{N_s}(G_s~L_s) \sum_{q=1}^{D} \bigg[3/4\sum_{k_2,m_2,n_2}\kappa_1^{(k_2)}\kappa_1^{(m_2)}\kappa_1^{(n_2)} P_{k_2,q} P_{ m_2,q} P_{n_2,p}X_{l_2,p}^{a}(k_2,m_2,n_2,q)+1/4\sum_{k_2,n_2}\kappa_2^{(k_2)} \kappa_1^{(n_2)} (P_{k_2,q}^2 P_{n_2,p}\\
& 5X_{l_2,p}^{b}(k_2,k_2,n_2,q)+P_{k_2,p} P_{k_2,q}P_{n_2,q}X_{l_2,p}^{c}(k_2,n_2,k_2,q))+1/4\sum_{n_2}\kappa_3^{(n_2)}  P_{n_2,q}^2P_{n_2,p} X_{l_2,p}^{d}(n_2,n_2,n_2,q)\bigg] +\sigma_{RxN}^2\bigg)\Bigg)/SNR^{req}_{i_2,p},
\end{aligned}\end{equation}}
\setcounter{equation}{\value{MYtempeqncnt}}
\hrulefill
\vspace*{4pt}
\end{figure*}

The Kerr nonlinearity is cubic, therefore, the product {\small $E[n_{\mathbf{i}}(f) n_{\mathbf{i}}^*(f)]$} depends on the product of six atoms. Note that only combinations with an equal number of conjugate/non-conjugate pairs are non-zero. In addition, it should be noted that {\small $\zeta_{\mathbf{i}}$} are independent and identically distributed random variables with zero mean and unit variance.
The considered CPE removes the average phase at the receiver ({\small $\phi$}) \cite{16}. In a perturbative frame, this corresponds to work with the following nonlinear interference noise

\small{\begin{equation}\label{eq:23}\begin{aligned}
n_{\mathbf{i}}^{'}=n_{\mathbf{i}}+j\phi \zeta_{\mathbf{i}}.
\end{aligned}\end{equation}}
The nonlinear noise variance in EGN model can be interpreted as a summation over the Second-Order Noise (SON) which is usually called the GN contribution, Fourth-Order Noise (FON), and Higher-Order Noise (HON) variances, $i.e.$
\small{\begin{equation}\label{eq:24}\begin{aligned}
\sigma_{EGN}^2=\sigma_{GN}^2+\sigma_{FON}^2+\sigma_{HON}^2.
\end{aligned}\end{equation}}

\newcounter{MYtempeqncnt21}
\begin{figure*}[!t]
\normalsize
\setcounter{MYtempeqncnt}{\value{equation}}
\setcounter{equation}{28}

\small{\begin{equation}\label{eq:29}\begin{aligned}
{}& \max \limits_{G_i,P_{i_2,p}}~\min\limits_{i_2,p,i}~ P_{i_2,p}\prod_{n=1}^{N_s}(G_n~L_n) / \bigg(\prod_{n=1}^{N_s}(G_n~L_n) (F (G_{BA}-1) h \nu B_{i_2})+\sum_{s=1}^{N_s}[(F(G_s-1)h\nu B_{i_2})\prod_{n=s+1}^{N_s}(G_n~L_n)] +\sum_{s=1}^{N_s}\prod_{n=1}^{s-1}(G_n~L_n)^3\\
&\prod_{n=s}^{N_s}(G_s~L_s) \sum_{q=1}^{D} \bigg[3/4\sum_{k_2,m_2,n_2}\kappa_1^{(k_2)}\kappa_1^{(m_2)}\kappa_1^{(n_2)} P_{k_2,q} P_{ m_2,q} P_{n_2,p}X_{l_2,p}^{a}(k_2,m_2,n_2,q)+1/4\sum_{k_2,n_2}\kappa_2^{(k_2)} \kappa_1^{(n_2)} (P_{k_2,q}^2 P_{n_2,p} \\
&5X_{l_2,p}^{b}(k_2,k_2,n_2,q)+P_{k_2,p} P_{k_2,q}P_{n_2,q} X_{l_2,p}^{c}(k_2,n_2,k_2,q))+1/4\sum_{n_2}\kappa_3^{(n_2)}  P_{n_2,q}^2P_{n_2,p} X_{l_2,p}^{d}(n_2,n_2,n_2,q)\bigg] +\sigma_{RxN}^2\bigg) \frac{1}{SNR^{req}_{i_2,p}}\\
&s.t. \left\{
    \begin{array}{ll}
P_{i_2,p,s}= P_{i_2,p,s-1}G_s~L_s\\
\sum_{i_2,p} P_{i_2,p}\prod_{n=1}^{s-1}(G_n~L_n) \leq P_{sat}^{MM-EDFA_s}\\
G_s \leq G_{max}^{MM-EDFA_s}
    \end{array}
\right.
\end{aligned}\end{equation}}
\setcounter{equation}{\value{MYtempeqncnt}}
\hrulefill
\vspace*{4pt}
\end{figure*}
\newcounter{MYtempeqncnt22}
\begin{figure*}[!t]
\normalsize
\setcounter{MYtempeqncnt}{\value{equation}}
\setcounter{equation}{29}

\small{\begin{equation}\label{eq:30}\begin{aligned}
{}&\min\limits_{P_{i_2,p},G_i} ~ \max\limits_{i_2,p,i}~SNR^{req}_{i_2,p} \bigg(\prod_{n=1}^{N_s}(G_n~L_n) (F (G_{BA}-1) h \nu B_{i_2})+\sum_{s=1}^{N_s}[(F(G_s-1)h\nu B_{i_2})\prod_{n=s+1}^{N_s}(G_n~L_n)] +\sum_{s=1}^{N_s}\prod_{n=1}^{s-1}(G_n~L_n)^3\prod_{n=s}^{N_s}(G_s~L_s) \\
&\sum_{q=1}^{D} \bigg[3/4\sum_{k_2,m_2,n_2}\kappa_1^{(k_2)}\kappa_1^{(m_2)}\kappa_1^{(n_2)} P_{k_2,q} P_{ m_2,q} P_{n_2,p}X_{l_2,p}^{a}(k_2,m_2,n_2,q)+1/4\sum_{k_2,n_2}\kappa_2^{(k_2)} \kappa_1^{(n_2)} (P_{k_2,q}^2 P_{n_2,p} 5X_{l_2,p}^{b}(k_2,k_2,n_2,q)\\
&+P_{k_2,p} P_{k_2,q}P_{n_2,q}X_{l_2,p}^{c}(k_2,n_2,k_2,q))+1/4\sum_{n_2}\kappa_3^{(n_2)}  P_{n_2,p}^2P_{n_2,q} X_{l_2,p}^{d}(n_2,n_2,n_2,q)\bigg] +\sigma_{RxN}^2\bigg) / \bigg(P_{i_2,p}\prod_{n=1}^{N_s}(G_n~L_n)\bigg)\\
& s.t. \left\{
    \begin{array}{ll}
P_{i_2,p,s}= P_{i_2,p,s-1}G_s~L_s\\
\sum_{i_2,p} P_{i_2,p}\prod_{n=1}^{s-1}(G_n~L_n) \leq P_{sat}^{MM-EDFA_s}\\
G_s \leq G_{max}^{MM-EDFA_s}
    \end{array}
\right.
\end{aligned}\end{equation}}
\setcounter{equation}{\value{MYtempeqncnt}}
\hrulefill
\vspace*{4pt}
\end{figure*}
It is shown in Appendix A that the nonlinear noise variance of the whole link becomes equal to (\ref{eq:25}).
Therefore, the nonlinear noise variance of the whole link becomes as (\ref{eq:25})
where {\small $P_{i_2,p}\triangleq P_{i_2,p,1}$} is the launched power at the first span, and {\small $P_{i_2,p,s}=P_{i_2,p,s-1} (G_s~L_s)$} is the launched power at the {\small $s$}th span. Note that the power at each span input is the multiplication of the power of the previous span by loss and MM-EDFA gain of that span.

\section{Problem statement}
The SNR margin available between the operating conditions and the error correction threshold allows for system aging, fiber repairs, and transient events. Extra margin ensures that service level agreements have been met for the lifetime of the system. The channel/mode with the lowest SNR margin is the most failure probability, thus minimum SNR margin maximization would minimize this probability. Based on the proposed notations in the previous section, the SNR of {\small $i_2$}th channel and {\small $p$}th mode can be expressed as \cite{19}
\addtocounter{equation}{1}
\small{\begin{equation}\label{eq:26}
SNR_{i_2,p} = \frac{P_{i_2,p}\prod_{n=1}^{N_s}(G_n~L_n)} {\sigma_{ASE}^2+\sigma_{EGN,i_2,p}^2+\sigma_{RxN}^2},
\end{equation}}
where {\small $\sigma_{RxN}^2$} is the receiver noise power. Moreover, the variance of ASE noise in the receiver can be expressed as
\small{\begin{equation}\label{eq:27}\begin{aligned}
{}&\sigma_{ASE}^2= \prod_{n=1}^{N_s}(G_n~L_n) (F (G_{BA}-1) h \nu B_{i_2})+\\
&\sum_{s=1}^{N_s}[(F(G_s-1)h\nu B_{i_2})\prod_{n=s+1}^{N_s}(G_n~L_n)],
\end{aligned}\end{equation}}
\newcounter{MYtempeqncnt23}
\begin{figure*}[!t]
\normalsize
\setcounter{MYtempeqncnt}{\value{equation}}

\setcounter{equation}{30}
\small{\begin{equation}\label{eq:31}\begin{aligned}
{}&\min\limits_{\hat{P}_{i_2,p},g_i} ~ \max\limits_{i_2,p,i}~log(SNR^{req}_{i_2,p})+log\bigg(\prod_{n=1}^{N_s}(e^{g_n}L_n) (F (G_{BA}-1) h \nu B_{i_2})+\sum_{s=1}^{N_s}[(F(e^{g_s}-1)h\nu B_{i_2})\prod_{n=s+1}^{N_s}(e^{g_n}L_n)]
 +\sum_{q=1}^{D}\sum_{k_2,m_2,n_2=1}^{N_{ch}}\\
& \kappa_1^{(k_2)}\kappa_1^{(m_2)}\kappa_1^{(n_2)} e^{\hat{P}_{k_2,p}+\hat{P}_{m_2, q}+\hat{P}_{n_2, q}}  3X_{l_2,p}^{a}(k_2,m_2,n_2,q) +\sum_{q=1}^{D}\sum_{k_2,n_2=1}^{ N_{ch}} \kappa_2^{(k_2)} \kappa_1^{(n_2)} (e^{2\hat{P}_{k_2,q}+\hat{P}_{n_2, p}} 5X_{l_2,p}^{b}(k_2,k_2,n_2,q)+e^{\hat{P}_{k_2,p}+\hat{P}_{k_2,q}+\hat{P}_{n_2,q}}\\
&X_{l_2,p}^{c}(k_2,n_2,k_2,q))+\sum_{q=1}^{D}\sum_{n_2=1}^{ N_{ch}} \kappa_3^{(n_2)} e^{2\hat{P}_{n_2,q}+\hat{P}_{n_2,p}} X_{l_2,p}^{d}(n_2,n_2,n_2,q)+\sigma_{RxN}^2
\bigg)-(\hat{P}_{i_2,p}+\sum_{n=1}^{N_s}g_n log(L_n))\\
&s.t. \left\{
    \begin{array}{ll}
\hat{P}_{i_2,p}^{s}= \hat{P}_{i_2,p}^{s-1}+g_s+log(L_s)\\
\sum_{i_2,p} e^{\hat{P}_{i_2,p}}\prod_{n=1}^{s-1}(e^{g_n}L_n) \leq P_{sat}^{MM-EDFA_s}\\
g_s \leq log(G_{max}^{MM-EDFA_s})
    \end{array}
\right.
\end {aligned}\end{equation}}
\setcounter{equation}{\value{MYtempeqncnt}}
\hrulefill
\vspace*{4pt}
\end{figure*}
\newcounter{MYtempeqncnt24}
\begin{figure*}[!t]
\normalsize
\setcounter{MYtempeqncnt}{\value{equation}}
\setcounter{equation}{31}

\small{\begin{equation}\label{eq:32}\begin{aligned}
{}&\min\limits_{\beta, \hat{P}_{i_2,p},g_i}~\beta\\
& s.t. \left\{
    \begin{array}{ll}
log(SNR^{req}_{i_2,p})+log\bigg(\prod_{n=1}^{N_s}(e^{g_n}L_n) (F (G_{BA}-1) h \nu B_{i_2})+\sum_{s=1}^{N_s}[(F(e^{g_s}-1)h\nu B_{i_2})\prod_{n=s+1}^{N_s}(e^{g_n}L_n)]
+ \sum_{q=1}^{ D} \sum_{k_2,m_2,n_2=1}^{ N_{ch}}\\
\kappa_1^{(k_2)}\kappa_1^{(m_2)}\kappa_1^{(n_2)}e^{\hat{P}_{k_2,p}+\hat{P}_{m_2, q}+\hat{P}_{n_2, q}} 3X_{l_2,p}^{a}(k_2,m_2,n_2,q)
+ \sum_{q=1}^{ D} \sum_{k_2,n_2=1}^{ N_{ch}} \kappa_2^{(k_2)}\kappa_1^{(n_2)}(e^{2\hat{P}_{k_2,q}+\hat{P}_{n_2,p}} 5X_{l_2,p}^{b}(k_2,k_2,n_2,q)\\
+e^{\hat{P}_{k_2,p}+\hat{P}_{k_2,q}+\hat{P}_{n_2,q}}X_{l_2,p}^{c}(k_2,n_2,k_2,q)) +\sum_{q=1}^{ D} \sum_{n_2=1}^{ N_{ch}} \kappa_3^{(n_2)}e^{2\hat{P}_{n_2,q}+\hat{P}_{n_2,p}} X_{l_2,p}^{d}(n_2,n_2,n_2,q)+\sigma_{RxN}^2\bigg)-(\hat{P}_{i_2,p}+\\
\sum_{n=1}^{N_s}g_n log(L_n)) \leq \beta\\
\hat{P}_{i_2,p}^{s}= \hat{P}_{i_2,p}^{s-1}+g_s+log(L_s)\\
\sum_{i_2,p} e^{\hat{P}_{i_2,p}}\prod_{n=1}^{s-1}(e^{g_n}L_n) \leq P_{sat}^{MM-EDFA_s}\\
g_s \leq log(G_{max}^{MM-EDFA_s})
    \end{array}
\right.
\end{aligned} \end{equation}}
\setcounter{equation}{\value{MYtempeqncnt}}
\hrulefill
\vspace*{4pt}
\end{figure*}
\newcounter{MYtempeqncnt25}
\begin{figure*}[!t]
\normalsize
\setcounter{MYtempeqncnt}{\value{equation}}
\setcounter{equation}{32}

\small{\begin{equation}\label{eq:33}\begin{aligned}
{}&\min\limits_{\beta, \hat{P}_{l},g_s}~\beta\\
&s.t. \left\{
    \begin{array}{ll}
\bigg[log(SNR^{req}_{l})+log\bigg(\prod_{n=1}^{N_s}(e^{g_n}L_n) (F (G_{BA}-1) h \nu B_{i_2})+\sum_{s=1}^{N_s}[(F(e^{g_s}-1)h\nu B_{i_2})\prod_{n=s+1}^{N_s}(e^{g_n}L_n)]
+ \sum_{l_1,l_2,l_3=1}^{D N_{ch}}\kappa_1^{{'}^{(l_1)}}\\
\kappa_1^{{'}^{(l_2)}}\kappa_1^{{'}^{(l_3)}} e^{\hat{P}_{l_1}+\hat{P}_{l_2}+\hat{P}_{l_3}} 3H_{l}^{a}(l_1,l_2,l_3)
+  \sum_{l_1,l_2,l_3=1}^{D N_{ch}}  \kappa_2^{{'}^{(l_1)}}\kappa_1^{{'}^{(l_2)}} (e^{2\hat{P}_{l_1}+\hat{P}_{l_2}}5H_{l}^{b}(l_1,l_1,l_2)+e^{\hat{P}_{l_1}+\hat{P}_{l_2}+\hat{P}_{l_3}}H_{l}^{c}(l_1,l_2,l_3))
\\
+ \sum_{l_1,l_2=1}^{D N_{ch}} \kappa_3^{{'}^{(l_1)}} e^{2\hat{P}_{l_1}+\hat{P}_{l_2}} H_{l}^{d}(l_1,l_1,l_2)+\sigma_{RxN}^2
\bigg)-(\hat{P}_{l}+\sum_{n=1}^{N_s}g_n log(L_n)) \bigg] \leq \beta\\
\hat{P}_{l}^{s}= \hat{P}_{l}^{s-1}+g_s+log(L_s)\\
\sum_{l} e^{\hat{P}_{l}}\prod_{n=1}^{s-1}(e^{g_n}L_n) \leq P_{sat}^{MM-EDFA_s}\\
g_s \leq log(G_{max}^{MM-EDFA_s})
    \end{array}
\right.
\end{aligned} \end{equation}}
\setcounter{equation}{\value{MYtempeqncnt}}
\hrulefill
\vspace*{4pt}
\end{figure*}
where {\small $F$} is the amplifier noise figure, {\small $G_{BA}$} is the booster amplifier gain, {\small $G_s$} is the MM-EDFA amplifier gain compensating the loss in each fiber span, {\small $h$} is Plank constant, and {\small $\nu$} is the central frequency. The SNR margin of {\small $i_2 $}th channel and {\small $p $}th mode can be defined as (\ref{eq:28})
where {\small $M_{i_2,p}$} denotes SNR margin of {\small $i_2$}th channel and {\small $p$}th mode, and {\small $SNR^{req}_{i_2,p}$} is the required SNR of {\small $i_2$}th channel and {\small $p$}th mode. Therefore, the minimum SNR margin maximization problem can be expressed as (\ref{eq:29})
where the second constraint means that the total power at the {\small $s$}-th MM-EDFA should be less than saturation power of the {\small $s$}-th MM-EDFA, and the third constraint means that {\small $s$}-th MM-EDFA gain should be less than the maximum possible gain. The optimization problem (\ref{eq:29}) is equivalent to the optimization problem (\ref{eq:30}), as the min-max of a function is equivalent to the max-min of its inverse.
(\ref{eq:30}) is a non-convex optimization problem. To solve this issue, by replacing {\small $P_{i_2,p}, G_i$} with {\small $e^{\hat{P}_{i_2,p}}, e^{g_i}$} in (\ref{eq:30}) noting that {\small $log(x)$} is a monotonic function in {\small $x$}, we get (\ref{eq:31}) with the same minimum as (\ref{eq:30}).
By defining the slack variable {\small $\beta$} (\ref{eq:31}) can be rewritten as (\ref{eq:32}).

We use the gradient descent algorithm in vector form to solve (\ref{eq:32}). This is performed by introducing a vector {\small$\mathbf{p}$} of dimension  {\small$DN_{ch}$} whose elements {\small$P_{l}; l=1,2,...,DN_{ch}$} are given by {\small$P_{n,m}$}, {\small$n=1,2,...,N_{ch},~m=1,2,...D$}.
In order to incorporate the values of {\small$B_n$}, we use a vector with the same dimension as {\small$\mathbf{p}$} defined as {\small$\mathbf{B} = [B_1, B_1,\ldots , B_1, B_2, B_2, \ldots , B_2,\ldots , B_{N_{ch}}, B_{N_{ch}}, \ldots, B_{N_{ch}}]$} in which each {\small$B_n$} has been repeated {\small$D$} times.
Also let {\small $X$} be a {\small $N_{ch} \times D \times N_{ch} \times N_{ch} \times N_{ch} \times D$} dimensional tensor with elements {\small $X_{l_2,p}^{(.)}(k_2,m_2,n_2,q)$}.
To match the latter dimensions with {\small$\mathbf{p}$}, we define a {\small $N_{ch}D \times N_{ch}D \times N_{ch}D \times N_{ch}D$} tensor, {\small $H$}, whose elements, {\small $H_{l}^{(.)}(l_1,l_2,l_3)$},
are equal to {\small $X_{l_2,p}^{(.)}(k_2,m_2,n_2,q)$} in different subscripts.
Thus, (\ref{eq:32}) can be expressed as (\ref{eq:33}).
(\ref{eq:33}) is a convex optimization problem (see Appendix B) and can be solved by many methods, e.g., Bisection method \cite{20}. The Bisection method converts the main problem into a feasibility problem by selecting a region and choosing a candidate for the objective function. The feasibility problem can be solved using the Lagrangian method \cite{21}.
In each step, the region boundaries are updated based on the obtained solution for the feasibility problem from the previous step. In this manner, the Bisection method is converged to the optimum objective.

\newcounter{MYtempeqncnt26}
\begin{figure*}[!t]
\normalsize
\setcounter{MYtempeqncnt}{\value{equation}}
\setcounter{equation}{33}

\small{\begin{equation}\label{eq:34}\begin{aligned}
{}&\beta+\sum_{l=1}^{DN_{ch}}\lambda_l \Bigg( \bigg[log(SNR^{req}_{l})+log\bigg(\prod_{n=1}^{N_s}(e^{g_n}L_n) (F (G_{BA}-1) h \nu B_{i_2})+\sum_{s=1}^{N_s}[(F(e^{g_s}-1)h\nu B_{i_2})\prod_{n=s+1}^{N_s}(e^{g_n}L_n)]
+ \sum_{l_1,l_2,l_3=1}^{D N_{ch}} \kappa_1^{{'}^{(l_1)}}\kappa_1^{{'}^{(l_2)}}\\
&\kappa_1^{{'}^{(l_3)}} e^{\hat{P}_{l_1}+\hat{P}_{l_2}+\hat{P}_{l_3}} 3H_{l}^{a}(l_1,l_2,l_3)
+  \sum_{l_1,l_2,l_3=1}^{D N_{ch}}  \kappa_2^{{'}^{(l_1)}}\kappa_1^{{'}^{(l_2)}} (e^{2\hat{P}_{l_1}+\hat{P}_{l_2}}5H_{l}^{b}(l_1,l_1,l_2)+e^{\hat{P}_{l_1}+\hat{P}_{l_2}+\hat{P}_{l_3}}H_{l}^{c}(l_1,l_2,l_3))
+\sum_{l_1,l_2=1}^{D N_{ch}} \kappa_3^{{'}^{(l_1)}}  e^{2\hat{P}_{l_1}+\hat{P}_{l_2}} \\
&H_{l}^{d}(l_1,l_1,l_2)+\sigma_{RxN}^2
\bigg)- (\hat{P}_{l}+\sum_{n=1}^{N_s}g_n log(L_n)) \bigg] - \beta \Bigg)+\sum_{s=1}^{N_s}\mu_s\bigg(\sum_{l} e^{\hat{P}_{l}}\prod_{n=1}^{s-1}(e^{g_n}L_n) - P_{sat}^{MM-EDFA_s}\bigg)+ \sum_{s=1}^{N_s} \nu_s  (g_s - \\
&log(G_{max}^{MM-EDFA_s}))
\end{aligned} \end{equation}}
\setcounter{equation}{\value{MYtempeqncnt}}
\hrulefill
\vspace*{4pt}
\end{figure*}
\newcounter{MYtempeqncnt27}
\begin{figure*}[!t]
\normalsize
\setcounter{MYtempeqncnt}{\value{equation}}
\setcounter{equation}{34}

\small{\begin{equation}\label{eq:35}\begin{aligned}
\inf\limits_{\hat{P}_{l},g_i}{} &\beta+\sum_{l=1}^{DN_{ch}}\lambda_l \Bigg( \bigg[log(SNR^{req}_{l})+log\bigg(\prod_{n=1}^{N_s}(e^{g_n}L_n) (F (G_{BA}-1) h \nu B_{i_2})+\sum_{s=1}^{N_s}[(F(e^{g_s}-1)h\nu B_{i_2})\prod_{n=s+1}^{N_s}(e^{g_n}L_n)]
+ \sum_{l_1,l_2,l_3=1}^{D N_{ch}} \kappa_1^{{'}^{(l_1)}}\\
&\kappa_1^{{'}^{(l_2)}}\kappa_1^{{'}^{(l_3)}} e^{\hat{P}_{l_1}+\hat{P}_{l_2}+\hat{P}_{l_3}} 3H_{l}^{a}(l_1,l_2,l_3)
+ \sum_{l_1,l_2,l_3=1}^{D N_{ch}}  \kappa_2^{{'}^{(l_1)}}\kappa_1^{{'}^{(l_2)}} (e^{2\hat{P}_{l_1}+\hat{P}_{l_2}}5H_{l}^{b}(l_1,l_1,l_2)+e^{\hat{P}_{l_1}+\hat{P}_{l_2}+\hat{P}_{l_3}}H_{l}^{c}(l_1,l_2,l_3))\\
&+  \sum_{l_1,l_2=1}^{D N_{ch}} \kappa_3^{{'}^{(l_1)}} e^{2\hat{P}_{l_1}+\hat{P}_{l_2}} H_{l}^{d}(l_1,l_1,l_2)+\sigma_{RxN}^2
\bigg)-(\hat{P}_{l}+\sum_{n=1}^{N_s}g_n log(L_n)) \bigg] - \beta \Bigg)+\sum_{s=1}^{N_s}\mu_s\bigg(\sum_{l} e^{\hat{P}_{l}}\prod_{n=1}^{s-1}(e^{g_n}L_n) - P_{sat}^{MM-EDFA_s}\bigg)\\
&+\sum_{s=1}^{N_s} \nu_s  (g_s - log(G_{max}^{MM-EDFA_s}))
\end{aligned} \end{equation}}
\setcounter{equation}{\value{MYtempeqncnt}}
\hrulefill
\vspace*{4pt}
\end{figure*}

\newcounter{MYtempeqncnt28}
\begin{figure*}[!t]
\normalsize
\setcounter{MYtempeqncnt}{\value{equation}}
\setcounter{equation}{35}

\small{\begin{equation}\label{eq:36}\begin{aligned}
{}&\Delta_{\lambda_l} = \bigg[log(SNR^{req}_{l})+log\bigg(\prod_{n=1}^{N_s}(e^{g_n}L_n) (F (G_{BA}-1) h \nu B_{i_2})+\sum_{s=1}^{N_s}[(F(e^{g_s}-1)h\nu B_{i_2})\prod_{n=s+1}^{N_s}(e^{g_n}L_n)]
+ \sum_{l_1,l_2,l_3=1}^{D N_{ch}} \kappa_1^{{'}^{(l_1)}}\kappa_1^{{'}^{(l_2)}}\kappa_1^{{'}^{(l_3)}}\\
& e^{\hat{P}_{l_1}+\hat{P}_{l_2}+\hat{P}_{l_3}} 3H_{l}^{a}(l_1,l_2,l_3)
+  \sum_{l_1,l_2,l_3=1}^{D N_{ch}}  \kappa_2^{{'}^{(l_1)}}\kappa_1^{{'}^{(l_2)}} (e^{2\hat{P}_{l_1}+\hat{P}_{l_2}}5H_{l}^{b}(l_1,l_1,l_2)+e^{\hat{P}_{l_1}+\hat{P}_{l_2}+\hat{P}_{l_3}}H_{l}^{c}(l_1,l_2,l_3))
+  \sum_{l_1,l_2=1}^{D N_{ch}} \kappa_3^{{'}^{(l_1)}} e^{2\hat{P}_{l_1}+\hat{P}_{l_2}} \\
&H_{l}^{d}(l_1,l_1,l_2)+\sigma_{RxN}^2
\bigg)- (\hat{P}_{l}+\sum_{n=1}^{N_s}g_n log(L_n)) \bigg] - \beta\\
&\Delta_{\mu_s} =\sum_{l} e^{\hat{P}_{l}}\prod_{n=1}^{s-1}(e^{g_n}L_n) - P_{sat}^{MM-EDFA_s}\\
&\Delta_{\nu_s} = g_s - log(G_{max}^{MM-EDFA_s})
\end{aligned} \end{equation}}
\setcounter{equation}{\value{MYtempeqncnt}}
\hrulefill
\vspace*{4pt}
\end{figure*}

\begin{algorithm}[tp!]
\caption{Bisection Method to Solve Convex Optimization Problem (\ref{eq:33}).}\label{alg:1}
\begin{algorithmic}[1]
\State \textbf{Initialization:} upper bound $u = 100$, and lower bound $l = -10$
\State $\beta \leftarrow u$
\State Solve convex problem (\ref{eq:33}) by Lagrangian method
\State \textbf{if} $\hat{\emph{P}}^{*(t)} == NAN$ \textbf{then} break
\State $\beta \leftarrow l$\;
\State Solve convex problem (\ref{eq:33}) by Lagrangian method\;
\State \textbf{if}$\hat{\emph{P}}^{*(t)} == NAN$ \textbf{then} break
\While{$u-l \leq \epsilon$}
\State $\beta \leftarrow (u + l)/2$
\State Solve convex problem (\ref{eq:33}) by Lagrangian method
\State \textbf{if} $\hat{\emph{P}}^{*(t)} == NAN$ \textbf{then} $l \leftarrow \beta$ \textbf{else} $u \leftarrow \beta$
\EndWhile
\end{algorithmic}
\end{algorithm}

Algorithm 1 summarizes the Bisection method for solving (\ref{eq:33}).
The first step in this algorithm is defining appropriate upper (\small{ $u$}) and lower (\small{ $l$}) bounds for the search region for \small{ $\beta$}. The defined upper bound is assigned to \small{ $\beta$}, and the problem (\ref{eq:30}) is converted to a feasibility problem. The feasibility problem is solved using the Lagrangian method. If the defined upper bound be lower than the optimal solution for \small{ $\beta$}, the feasibility problem would not have a solution. In other words, the feasible set is empty and a higher upper bound should be used. The defined lower bound can be tested and adjusted in the same manner.

In the second step, the upper bound \small{ $(u+l)/2$} is assigned to \small{ $\beta$}. The same as the first step, it is tested whether \small{ $(u+l)/2$} is the upper or lower bound of the feasible set. Thereby the upper and lower bounds are updated. The second step is repeated until convergence.

At each iteration of Algorithm 1, the feasibility problem is solved by the Lagrange duality method as summarized in Algorithm 2. Furthermore, the second constraint can be relaxed since it is satisfied by the objective function. The Lagrangian function of (\ref{eq:33}) is given by (\ref{eq:34}) where \small{ $\lambda_l, \mu_s$}, and \small{ $\nu_s \in R^{+}$} are the Lagrangian multipliers. Accordingly, the Lagrangian dual function of (\ref{eq:33}) can be expressed as (\ref{eq:35}).

(\ref{eq:35}) is a convex problem with respect to \small{ $\hat{P}_{l}, g_i$}, since the dual problem is a convex optimization problem \cite{21}. Note that at each iteration of Algorithm 2, \small{ $\lambda_l, \mu_s$}, and \small{ $ \nu_s$} are updated based on the derivative of (\ref{eq:35}) with respect to \small{ $\lambda_l, \mu_s$}, and \small{ $ \nu_s $} which are shown in (\ref{eq:36}).
\begin{algorithm}[bp!]
\caption{Lagrangian Duality Method to Solve the Convex Problem (\ref{eq:35}).}\label{alg:2}
\begin{algorithmic}[1]
\State \textbf{Initialization:} iteration counter $t=0$, step size parameter $a>0, b>0, c>0$, and $\lambda^{(0)} \succeq 0, \mu^{(0)} \succeq 0, \nu^{(0)} \succeq 0$
\While{achieving convergence}
\State Solve convex problem (\ref{eq:35}) with fixed $\lambda, \mu, and \nu$, and obtain optimal power $\hat{\emph{P}}^{*(t)}$, and optimal gain $\emph{g}^{*(t)}$
\State $\lambda^{(t+1)} =\bigg[ \lambda^{(t)} –a \bigg(\Delta_{\lambda}\bigg) \bigg]^{+} $
\State $\mu^{(t+1)} =\bigg[ \mu^{(t)} –b \bigg(\Delta_{\mu}\bigg) \bigg]^{+} $
\State $\nu^{(t+1)} =\bigg[ \nu^{(t)} –c \bigg(\Delta_{\nu}\bigg) \bigg]^{+} $
\State \textbf{update} $t=t+1$
\EndWhile
\end{algorithmic}
\end{algorithm}

\begin{figure}[htp]
  \centering
  \subfigure[]{\includegraphics[scale=0.25]{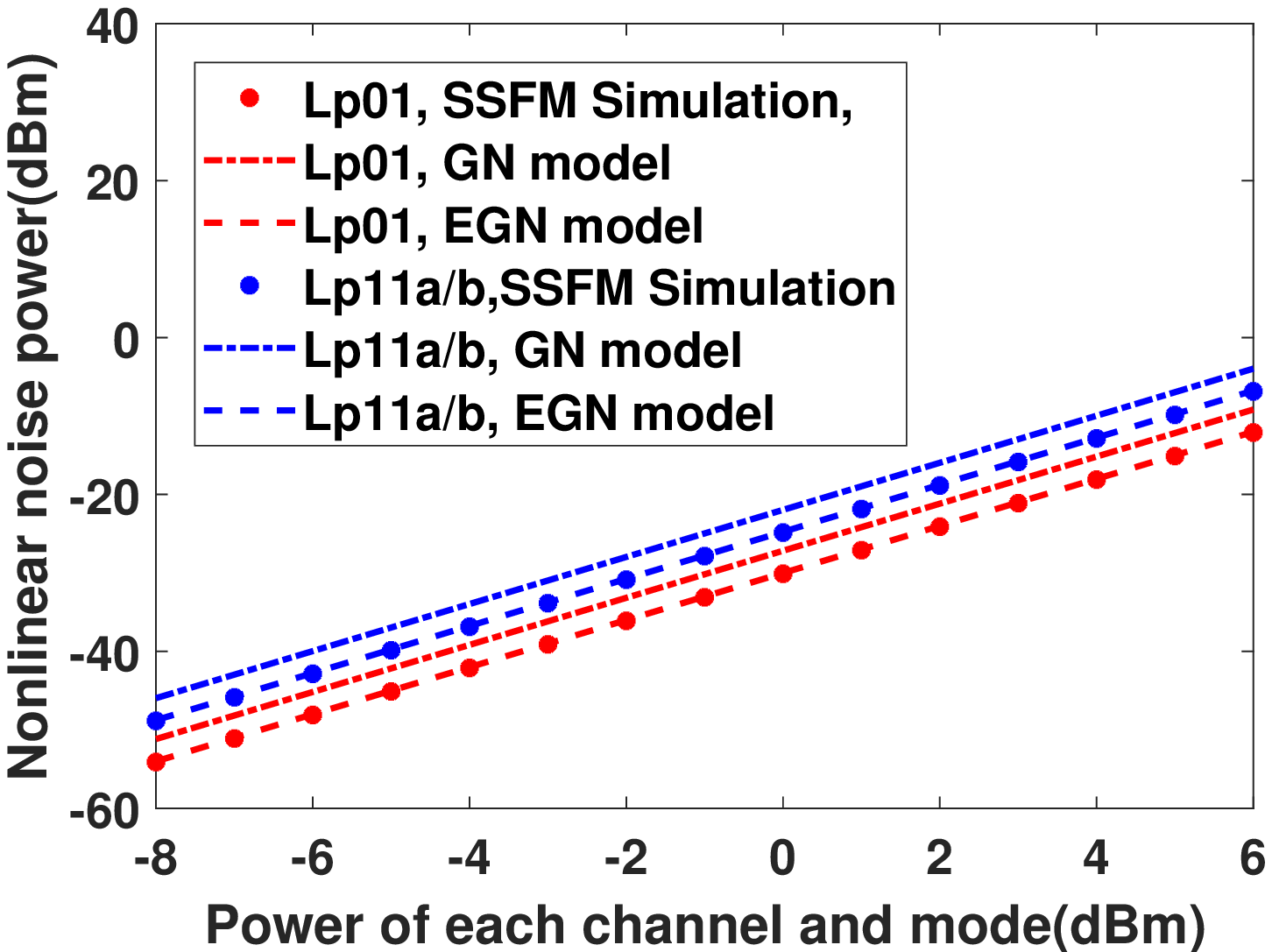}\label{fig:2a}}\quad
  \subfigure[]{\includegraphics[scale=0.25]{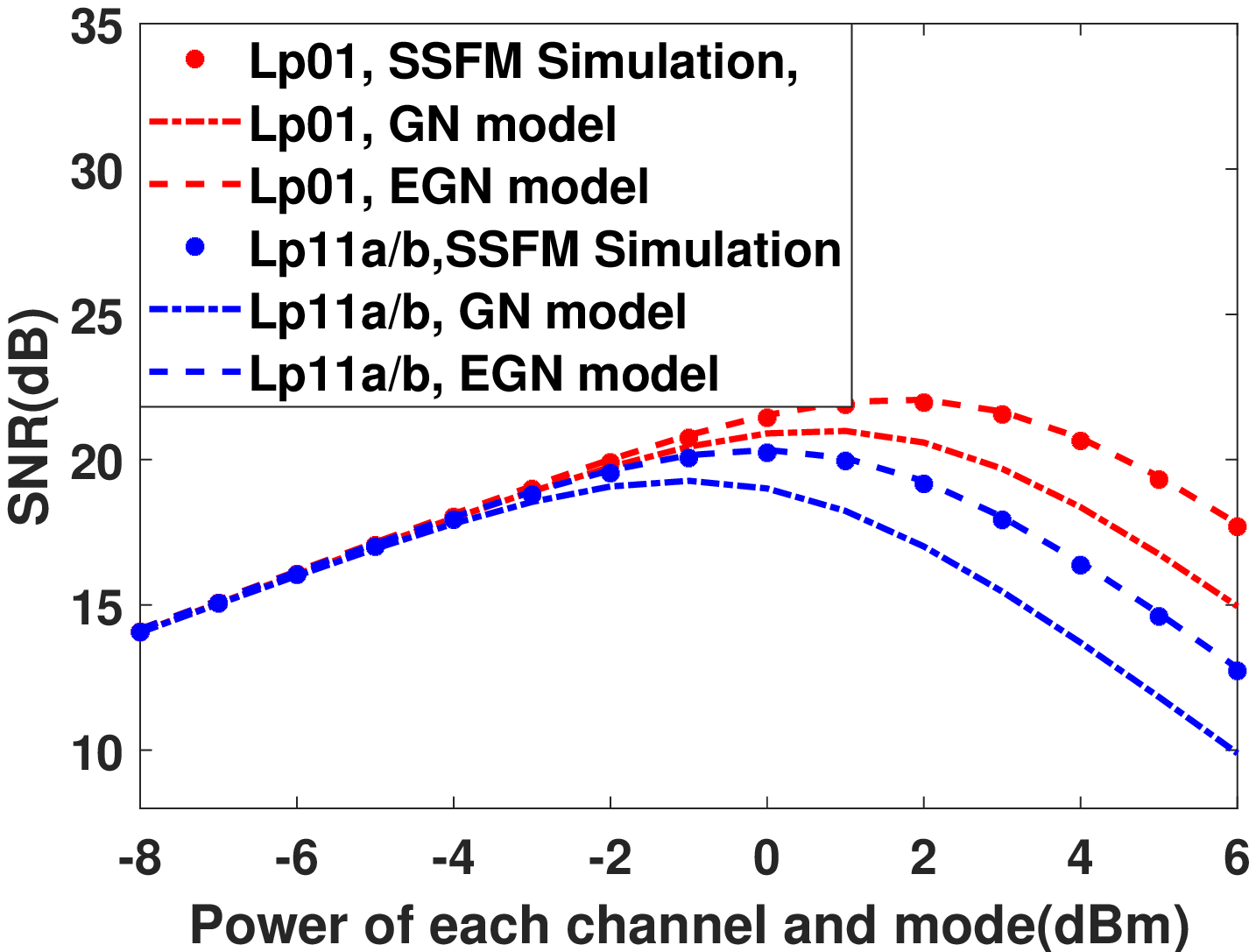}\label{fig:2b}}
\caption{a) Nonlinear noise power and b) SNR based on the proposed EGN model and the SSFM simulation versus launched power of each channel and mode for MDM-WDM ({\small $D=3, N_{ch}=3$}) system.}
    \label{fig:2}
\end{figure}
\textit{Remark 1}: In Algorithm 2, (\ref{eq:35}) is solved at each iteration as a function of {\small$\hat{P}_l$} using the gradient descent algorithm which will converge to its optimum solution due to the convexity of the problem. This procedure is repeated by Algorithm 1 in the "While loop", by which the minimum SNR margin is improved successively until convergence to the maximum value.
Note that Algorithm 1 will stop searching while the difference between upper and lower bound becomes less than \small{ $\epsilon$}.
\section{Simulation results}
\subsection{Accuracy of the proposed EGN model formulation}

\begingroup
\tabcolsep = 8.0pt
\def\arraystretch{1}
 \begin{table}[bp]
 \centering
   \caption{Simulation parameters.}
   \begin{tabular}{| c | c | c |}
    \hline
     {\footnotesize Coefficient} &	{\footnotesize Symbol} & {\footnotesize Value}\\ \hline
     {\footnotesize Symbol rate} & {\footnotesize $R_n$} & {\footnotesize $32~GBaud$}\\ \hline
     {\footnotesize Channel Spacing} & {\footnotesize $B_w$} & {\footnotesize $50~GHz$}\\ \hline
     {\footnotesize Channel bandwidth} & {\footnotesize $B_n$} & {\footnotesize $32~GHz$}\\ \hline
     {\footnotesize Span length} & {\footnotesize $L_s$} & {\footnotesize $80~km$}\\ \hline
     {\footnotesize Nonlinearity coefficient}& {\footnotesize $\gamma$} & {\footnotesize $1.3~1/(W.km)$}\\ \hline
     {\footnotesize Center frequency}& {\footnotesize $\nu$} & {\footnotesize $1550~nm$}\\ \hline
     {\footnotesize Noise figure}& {\footnotesize $F$} & {\footnotesize $6~dB$}\\ \hline
     {\footnotesize MM-EDFA maximum gain}& {\footnotesize $G_{max}^{EDFA}$} & {\footnotesize $30~dB$}\\ \hline
     {\footnotesize MM-EDFA saturation power}& {\footnotesize $P_{sat}^{EDFA}$} & {\footnotesize $25~dBm$}\\ \hline
     {\footnotesize Receiver noise power}& {\footnotesize $\sigma_{Rxn}^2$} & {\footnotesize $-28~dBm$}\\ \hline
     {\footnotesize Booster amplifier gain}& {\footnotesize $G_{BA}$} & {\footnotesize $20~dB$}\\ \hline
     \end{tabular}
    \label{table:2}
\end{table}
\endgroup

\begingroup
\tabcolsep = 6.0pt
\def\arraystretch{1}
 \begin{table}[bp]
 \centering
   \caption{Nonlinear coupling coefficient $(f_{pq})$ \cite{9}.}
   \begin{tabular}{| c | c | c | c | c | c | c |}
    \hline
     {\footnotesize mq} & {\footnotesize LP01} & {\footnotesize LP11a} & {\footnotesize LP11b}  & {\footnotesize LP02}  & {\footnotesize LP21a}  & {\footnotesize LP21b}\\ \hline
     {\footnotesize LP01} &\footnotesize{ 1}&\footnotesize{ 0.661}&\footnotesize{ 0.661}&\footnotesize{0.734 }&\footnotesize{0.455}&\footnotesize{ 0.455}\\ \hline
     {\footnotesize LP11a} &\footnotesize{ 0.660 }&\footnotesize{1.053}&\footnotesize{ 1.053 }&\footnotesize{0.369}&\footnotesize{ 0.608 }&\footnotesize{0.608}\\ \hline
     {\footnotesize LP11b} &\footnotesize{ 0.660}&\footnotesize{ 1.053 }&\footnotesize{1.053 }&\footnotesize{0.369}&\footnotesize{ 0.608}&\footnotesize{ 0.608}\\ \hline
     {\footnotesize LP02} &\footnotesize{ 0.731 }&\footnotesize{0.369}&\footnotesize{ 0.369}&\footnotesize{ 0.964 }&\footnotesize{0.335}&\footnotesize{ 0.335}\\ \hline
     {\footnotesize LP21a} &\footnotesize{ 0.455}&\footnotesize{ 0.608 }&\footnotesize{0.608}&\footnotesize{ 0.335}&\footnotesize{ 0.930 }&\footnotesize{0.930}\\ \hline
     {\footnotesize LP21b} &\footnotesize{ 0.455 }&\footnotesize{0.608}&\footnotesize{ 0.608 }&\footnotesize{0.335}&\footnotesize{ 0.930}&\footnotesize{ 0.930}\\ \hline
    \end{tabular}
    \label{table:3}
\end{table}
\endgroup
\begingroup
\tabcolsep = 6.0pt
\def\arraystretch{1}
 \begin{table}[bp]
 \centering
   \caption{Attenuation ($\alpha_p$ ($dB/km$)), and dispersion terms ($\beta_{1_p}$($ns/km$), $\beta_{2_p}$ ($ps^2/km$), and $\beta_{3_p}$ ($ps^3/km$)) \cite{9}.}
   \begin{tabular}{| c | c | c | c | c | c |c |}
    \hline
        & {\footnotesize LP01} & {\footnotesize LP11a} & {\footnotesize LP11b} & {\footnotesize LP02} & {\footnotesize LP21a} & {\footnotesize LP21b} \\ \hline
     {\footnotesize $\alpha_p$} & {\footnotesize $0.226$} & {\footnotesize $0.226$} & {\footnotesize $0.226$} & {\footnotesize $0.226$} & {\footnotesize $0.226$} & {\footnotesize $0.226$}\\ \hline
     {\footnotesize $\beta_{1_p}$} &\footnotesize{ 0}&\footnotesize{ 6.5 }&\footnotesize{ 6.5}&{\footnotesize 9.9}&\footnotesize{ 12}&\footnotesize{ 12}\\ \hline
     {\footnotesize $\beta_{2_p}$} &\footnotesize{ 31.86 }&\footnotesize{ 34.8 }&\footnotesize{ 34.8}&\footnotesize{2.93}&\footnotesize{26.51 }&\footnotesize{26.51}\\ \hline
     {\footnotesize $\beta_{3_p}$} &\footnotesize{ 0.1452 }&\footnotesize{0.1452 }&\footnotesize{0.1452 }&\footnotesize{ 0.1452 }&\footnotesize{ 0.1452 }&\footnotesize{0.1452}\\ \hline
\end{tabular}
    \label{table:4}
\end{table}
\endgroup
\begingroup
\tabcolsep = 8.0pt
\def\arraystretch{1}
\begin{table}[tp!]
 \centering
\caption{Lightpath number propagated by each channel and mode.}
   \begin{tabular}{|c|c|c|c|}
    \hline
     \multicolumn{2}{|c|}{ \textcolor{blue}{\textbf{SMF-WDM}}}&\multicolumn{2}{c|}{ \textcolor{blue}{\textbf{MDM-single channel}}}\\
      \hline
      {\footnotesize	 \textcolor{red}{Channel}} & {\footnotesize	 \textcolor{red}{lightpath \#}} & {\footnotesize	 \textcolor{red}{Mode}} & {\footnotesize	 \textcolor{red}{lightpath \#}} \\ \hline
      {\footnotesize	 $1,2$} & {\footnotesize	 $L1$} & {\footnotesize	 LP01} & {\footnotesize	 $L1$} \\ \hline
      {\footnotesize	 $3,4$} & {\footnotesize	 $L2$} & {\footnotesize	 Lp11a} & {\footnotesize	 $L2$}  \\ \hline
      {\footnotesize	 $5,6$} & {\footnotesize	 $L3$} & {\footnotesize	 Lp11b} & {\footnotesize	 $L3$}  \\ \hline
      {\footnotesize	 $7,8$} & {\footnotesize	 $L4$} & {\footnotesize	 Lp02} & {\footnotesize	 $L4$} \\ \hline
      {\footnotesize	 $9,10$} & {\footnotesize	 $L5$} & {\footnotesize	 Lp21a} & {\footnotesize	 $L5$} \\ \hline
      {\footnotesize	 $11$} & {\footnotesize	 $L6$} & {\footnotesize	 Lp21b} & {\footnotesize	 $L6$} \\ \hline
    \end{tabular}
    \label{table:5}
\end{table}
\endgroup
\begin{figure}[tp!]
    \centering
\includegraphics[width=0.5\linewidth, height=3cm]{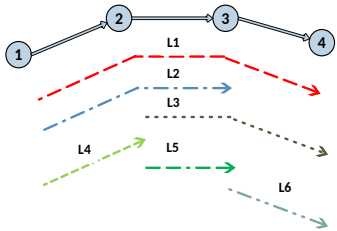}
\caption{A $4$-node linear network with $6$ lightpaths.}
\label{fig:3}
\end{figure}

In this section, the signal propagation is simulated by approximating the output of the Manakov equation using the well-known SSFM method with logarithmic step-size \cite{22} in the Python/Tensorflow environment. The simulation parameters and their values are presented in Tables \ref{table:2}, \ref{table:3}, and \ref{table:4}.

Figs 2a and 2b respectively show nonlinear noise power and SNR based on the proposed EGN model and the SSFM simulation versus launched power of each channel and mode for MDM-WDM (\small{ $D=3, N_{ch}=3$}) system. Note that only the SNR of the central channel is plotted, and QPSK modulation is considered in SSFM.
As seen in Fig. 2a, the proposed EGN model and SSFM simulation are in close agreement in all power ranges. However, the GN model overestimates the nonlinear noise power of SSFM simulation with QPSK modulation, since the nonlinear noise power of the Gaussian constellation is higher than QPSK.
As seen in Fig. 2b, a close agreement can be seen between the proposed EGN model, GN model, and SSFM simulation in the linear region, since the linear effects are dominant in this region. However, a fixed gap appears between the GN model with the proposed EGN mode and SSFM simulation in the nonlinear region.

\subsection{Minimum SNR margin maximization}
In this section the joint optimized power and gain allocation is performed based on minimum SNR margin maximization. Three scenarios are considered including a) best equal power, b) optimized power, and c) joint optimized power and gain.
In the first scenario, equal powers are considered for different channels and modes with equal MM-EDFA gain in all spans. It is worth mentioning that the MM-EDFA gain is equal to span loss.
In the second scenario, different powers are allocated to different channels and modes with equal MM-EDFA gain in all spans.
In the third case, allocated powers to each channel and mode are different. Moreover, the MM-EDFA gain for each span is optimized separately.
The SMF-WDM (\small{ $D=1, N_{ch}=11$}), and MDM-single channel (\small{ $D=6, N_{ch}=1$}) systems are considered \cite{3}, \cite{17}.

The point-to-point links often have a homogeneous set where different channels/modes experience the same interacting channels/modes. However, in multi-node linear networks, channel/modes may propagate different distances thus accumulate different nonlinear noise, experience fragmentation/partial utilization thereby see different interacting channels/modes and observe different MM-EDFA gains.
The \small{ $4$}-node linear network \cite{24} with \small{ $6$} lightpaths shown in Fig. 3 is considered in this paper for joint power and gain allocation. The lightpath number propagated by each channel and mode is presented in Table \ref{table:5}. Note that the BPSK modulation with \small{ $5.5~dB$} required SNR is considered \cite{23}.

\begin{figure}[tp]
  \centering
  \subfigure[]{\includegraphics[scale=0.25]{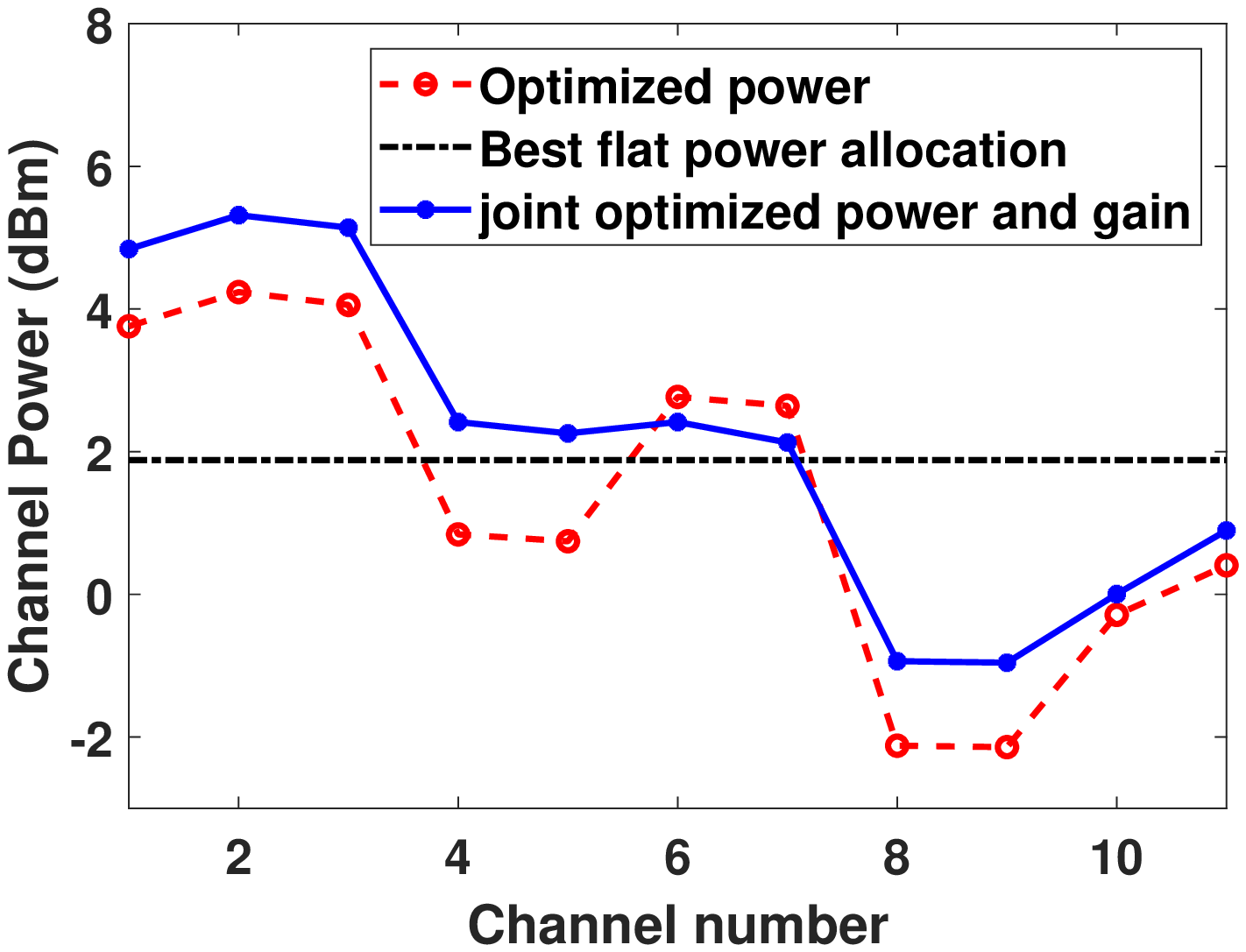}\label{fig:4a}}\quad
  \subfigure[]{\includegraphics[scale=0.25]{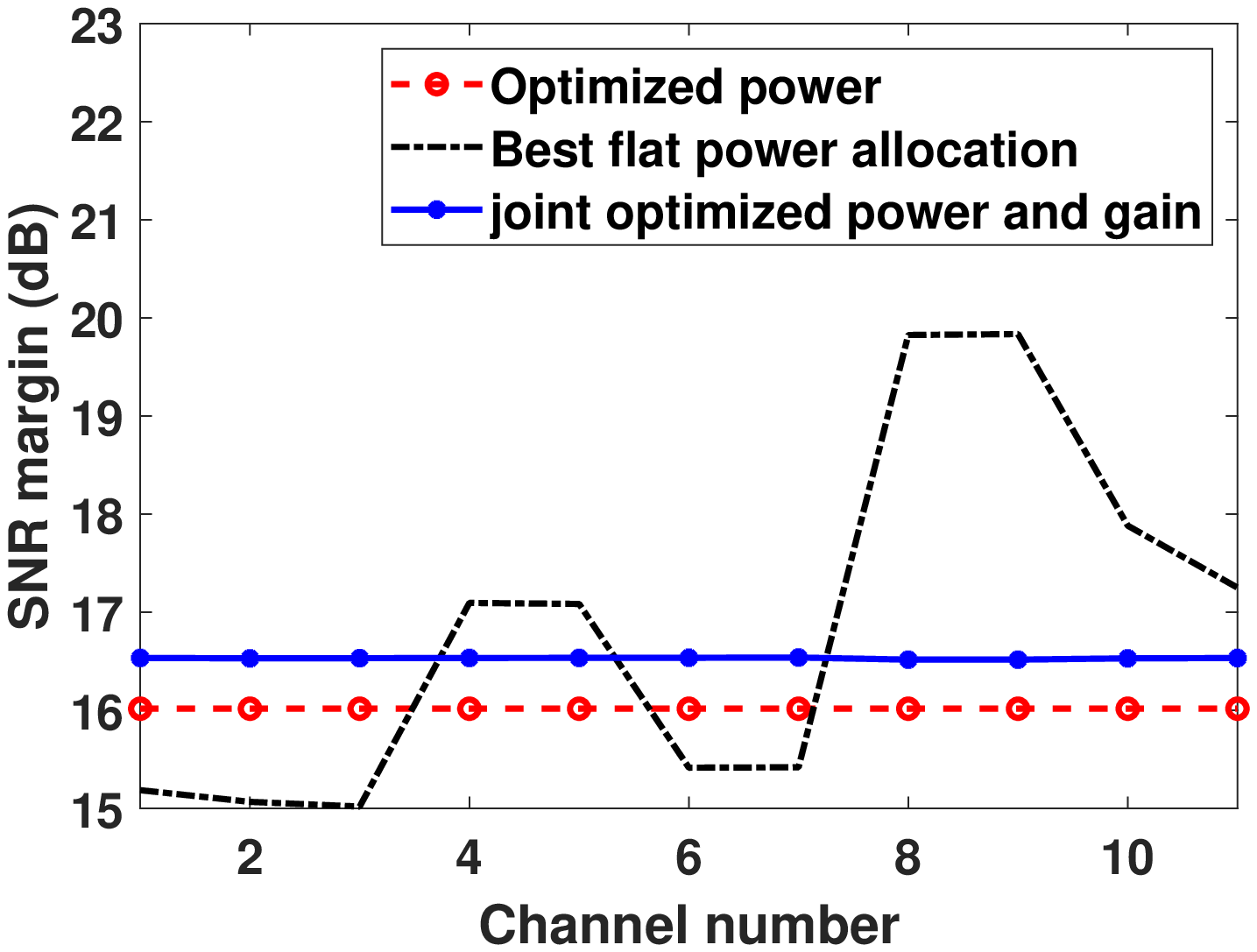}\label{fig:4b}}
    \subfigure[]{\includegraphics[scale=0.25]{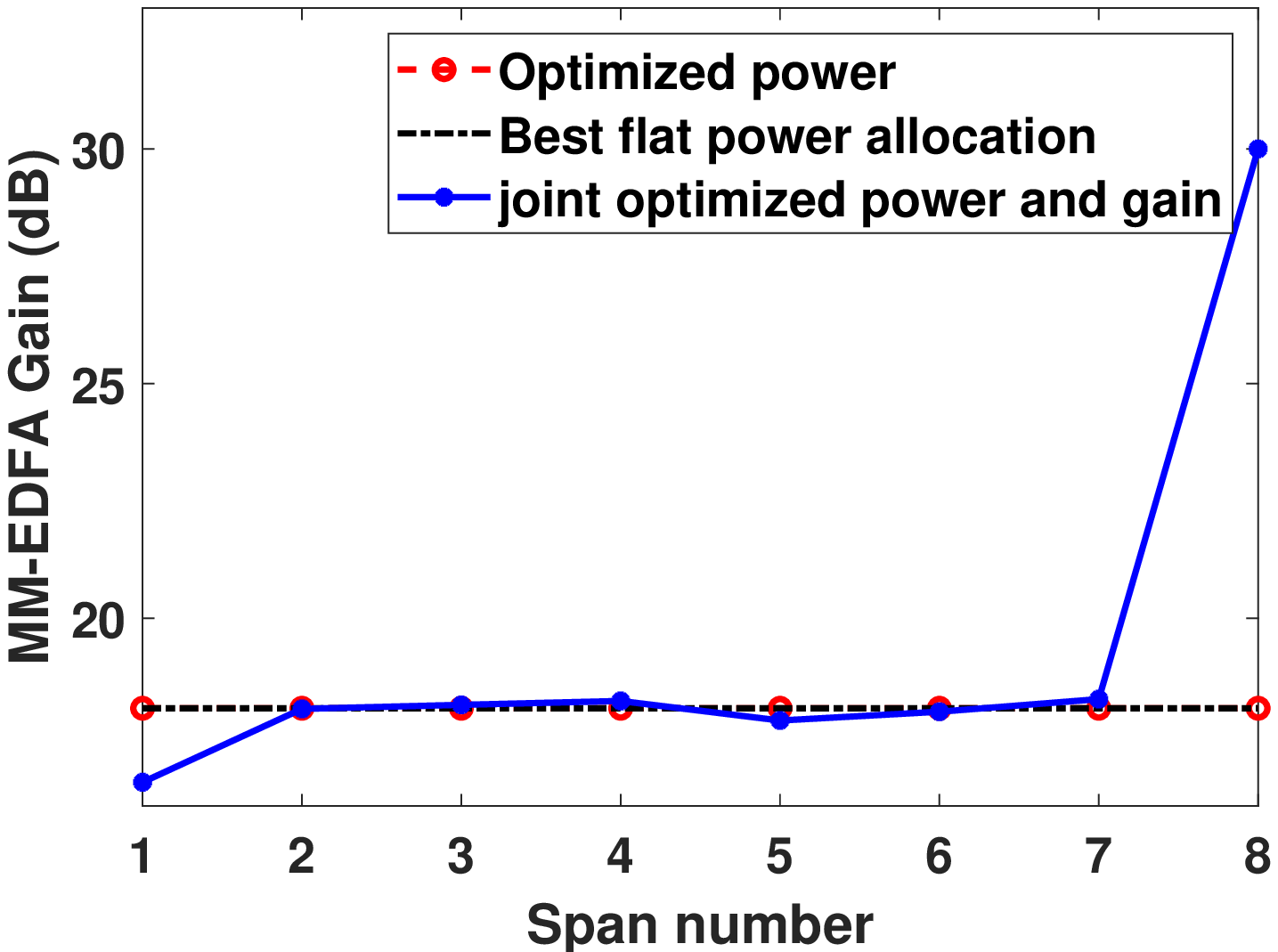}\label{fig:4c}}
\caption{a) Channel power and b) SNR margin versus channel number, and c) MM-EDFA gain versus span number, for joint optimized power and gain allocation, optimized power allocation and best equal power allocation, considering SMF-WDM system.}
    \label{fig:4}
\end{figure}

Figs 4a and 4b respectively depict channel power and SNR margin versus channel number, and Fig 4c shows MM-EDFA gain versus span number, for joint optimized power and gain allocation, optimized power allocation, and best equal power allocation, considering SMF-WDM system.
The obtained minimum SNR margins are \small{ $14.89~dB$} for best equal power allocation, \small{ $15.89~dB$} for optimized power allocation, and \small{ $16.51~dB$} for joint optimized power and gain allocation. Therefore, the obtained improvements of joint optimized power and gain allocation over optimized power allocation and best equal power allocation are \small{ $0.62$} and \small{ $1.62~dB$}, respectively.
The joint power and amplifier gain allocation observes higher degrees of freedom and thus obtains more improvements over optimized power allocation and best equal power allocation.
Central channel indices observe higher nonlinear noise (lower SNR margins). Therefore, they should be allocated higher power to have a reliable link.
Moreover, propagating channels in longer lightpath should be allocated higher power, since the longer the lightpath propagated by a channel, the more ASE/nonlinear noise power is added to that channel.
In joint optimized power and gain, the last MM-EDFA gain is set to its maximum possible value, this result can be deduced from SNR formulation where all terms except the receiver noise are scaled with {\small$G_{N_s}$}, and to minimize the contribution of the receiver noise term, the maximum possible value should be chosen for {\small$G_{N_s}$}.
In joint optimized power and gain, the last MM-EDFA gain is set to its maximum possible value, this result can be deduced from SNR formulation where all terms except the receiver noise are scaled with {\small$G_{N_s}$}, and to minimize the contribution of the receiver noise term, the maximum possible value should be chosen for {\small$G_{N_s}$}.

\begin{figure}[tp]
  \centering
  \subfigure[]{\includegraphics[scale=0.25]{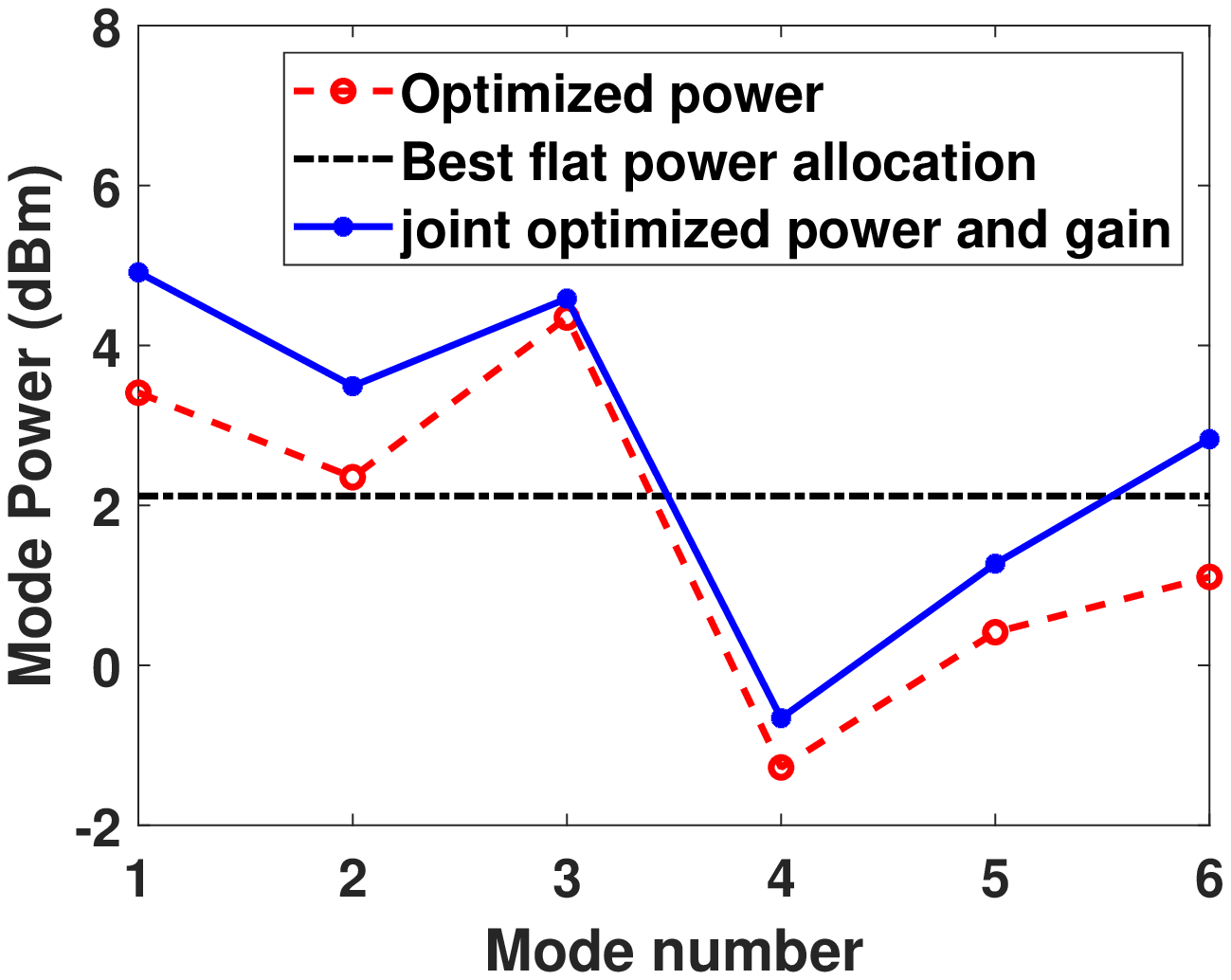}\label{fig:5a}}\quad
  \subfigure[]{\includegraphics[scale=0.25]{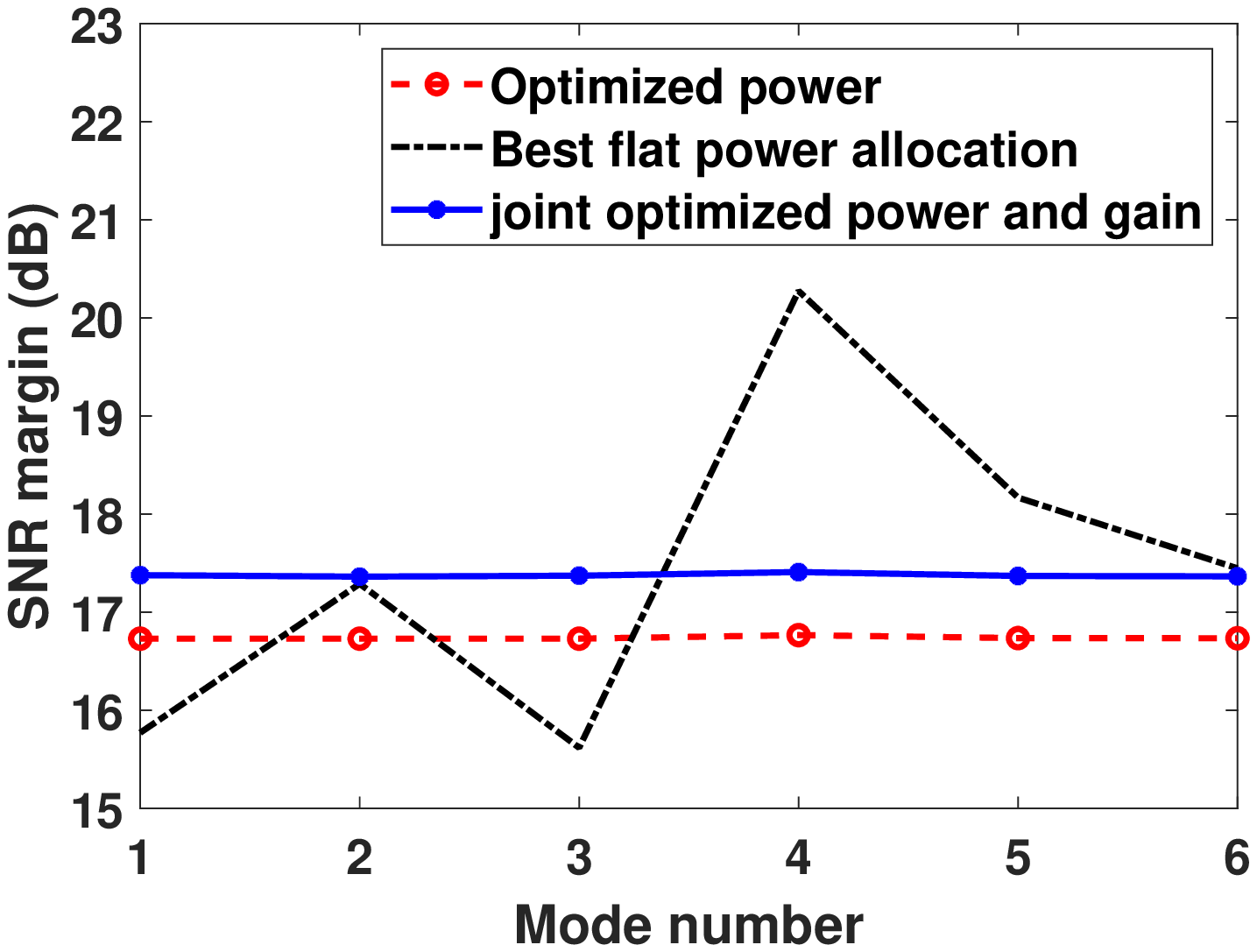}\label{fig:5b}}
  \subfigure[]{\includegraphics[scale=0.25]{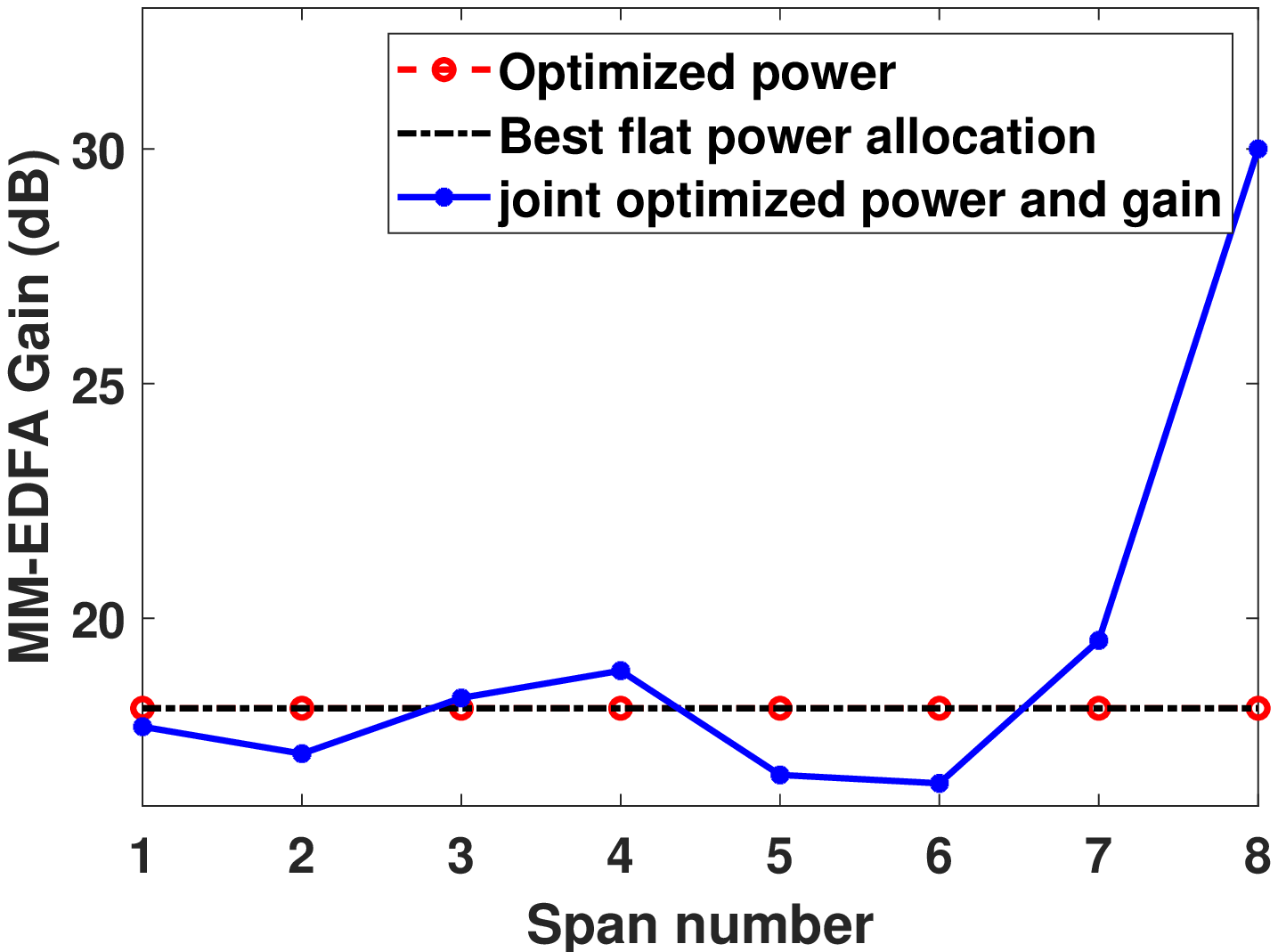}\label{fig:5c}}
\caption{a) Mode power and b) SNR margin versus mode number, and c) MM-EDFA gain versus span number, for joint optimized power and gain allocation, optimized power allocation and best equal power allocation considering MDM-single channel system.}
    \label{fig:5}
\end{figure}

Figs 5a and 5b respectively depict mode power and SNR margin versus mode number, and Fig 5c shows MM-EDFA gain versus span number, for joint optimized power and gain allocation, optimized power allocation, and best equal power allocation considering MDM-single channel system.
The obtained minimum SNR margins are {\small$15.63~dB$} for best equal power allocation, {\small$16.74~dB$} for optimized power allocation, and {\small$17.35~dB$} for joint optimized power and gain allocation. Therefore, the obtained improvements of joint optimized power and gain allocation over optimized power allocation and best equal power allocation are {\small$0.61$} and {\small$1.72~dB$}, respectively.
The main difference between different modes of the same channel is their spatial profiles. The LP11a/b mode has a larger spatial profile and therefore, a higher overlap with the other modes. Therefore, this mode has more nonlinear noise power and lower SNR margin. Accordingly, it should be allocated higher power than the other modes.
Neither the allocated powers nor the MM-EDFA gains are not symmetric, since different modes have different nonlinear noise power which is not symmetric due to the nonlinear coupling.

\section{Conclusion}
Achieving reliable communication over different channels and modes is one of the main goals in MDM-WDM networks, and is generally quantified through minimum SNR margin. In this paper, the EGN model formulation is derived for MDM-WDM systems for the first time and verified through SSFM simulations. Based on the proposed EGN model, the joint optimized power and MM-EDMA gain is proposed considering the minimum SNR margin maximization over different channels and modes in a multi-node linear network. It is shown that that the joint optimized power and MM-EDMA gain allocation results in minimum SNR margin improvement compared to optimized power allocation and the best equal power allocation. For instance, considering MDM-single channel and single mode fiber-WDM systems, joint power and gain optimization improves the minimum SNR margin {\small$0.61~dB$} and {\small$1.72~dB$} compared to optimized power allocation, respectively. Moreover, in these equations we have

\appendix
\section*{Appendix A: The nonlinear noise variance of $i_2$th channel and $p$th mode}
The GN, FON, and HON terms in EGN model can be expressed as (\ref{eq:37}), (\ref{eq:38}), and (\ref{eq:39}), respectively.
\newcounter{MYtempeqncnt6}
\begin{figure*}[!t]
\normalsize
\setcounter{MYtempeqncnt}{\value{equation}}
\setcounter{equation}{36}
\footnotesize{\begin{equation}\label{eq:37}\begin{aligned}
{}&\sigma_{GN,i_2,i_3}^2=\sum_{{\mathbf{k,m,n}}}\kappa_1^{({\mathbf{k}})} \kappa_1^{({\mathbf{m}})} \kappa_1^{({\mathbf{n}})}\Bigg(\iiint_{-\infty}^{\infty} \braket{\eta(f,f_1,f_2)|\eta(f,f_1,f_2)} \tilde{W}_{Tx}^{\mathbf{k}*}(f+f_1+f_2) \tilde{W}_{Tx}^{\mathbf{m}}(f+f_2) g^{\mathbf{i}*}(f) \tilde{W}_{Tx}^{\mathbf{n}}(f+f_1) \braket{k_3|m_3}\braket{i_3|n_3} \\
& \tilde{W}_{Tx}^{\mathbf{m}*}(f+f_2)\tilde{W}_{Tx}^{\mathbf{k}}(f+f_1+f_2) \tilde{W}_{Tx}^{\mathbf{n}*}(f+f_1) g^{\mathbf{i}}(f) \braket{m_3|k_3}\braket{n_3|i_3}df_1  df_2 df+\iiint_{-\infty}^{\infty} \braket{\eta(f,f_1,f_2)|\eta(f,f_1,f_2)} \tilde{W}_{Tx}^{\mathbf{k}*}(f+f_1+f_2) \tilde{W}_{Tx}^{\mathbf{m}}(f+f_2)\\
& g^{\mathbf{i}*}(f) \tilde{W}_{Tx}^{\mathbf{n}}(f+f_1) \braket{k_3|m_3} \braket{i_3|n_3}\tilde{W}_{Tx}^{\mathbf{n}*}(f+f_2) \tilde{W}_{Tx}^{\mathbf{k}}(f+f_1+f_2) \tilde{W}_{Tx}^{\mathbf{m}*}(f+f_1) g^{\mathbf{i}}(f) \braket{n_3|k_3}\braket{m_3|i_3}df_1  df_2 df\Bigg)+\Bigg(\iiint_{-\infty}^{\infty}\\
& \braket{\eta(f,f_1,f_2)|\eta(f,f_1,f_2)} \tilde{W}_{Tx}^{\mathbf{k}*}(f+f_1+f_2) \tilde{W}_{Tx}^{\mathbf{n}}(f+f_2) g^{\mathbf{i}*}(f) \tilde{W}_{Tx}^{\mathbf{k}}(f+f_1) \braket{k_3|n_3}\braket{i_3|k_3} \tilde{W}_{Tx}^{\mathbf{m}*}(f+f_2) \tilde{W}_{Tx}^{\mathbf{m}}(f+f_1+f_2) \tilde{W}_{Tx}^{\mathbf{n}*}(f+f_1) g^{\mathbf{i}}(f)\\
& \braket{m_3|m_3}\braket{n_3|i_3}df_1  df_2 df+\iiint_{-\infty}^{\infty} \braket{\eta(f,f_1,f_2)|\eta(f,f_1,f_2)} \tilde{W}_{Tx}^{\mathbf{k}*}(f+f_1+f_2) \tilde{W}_{Tx}^{\mathbf{n}}(f+f_2) g^{\mathbf{i}*}(f) \tilde{W}_{Tx}^{\mathbf{k}}(f+f_1) \braket{k_3|n_3}\braket{i_3|k_3} \tilde{W}_{Tx}^{\mathbf{n}*}(f+f_2)\\
&  \tilde{W}_{Tx}^{\mathbf{m}}(f+f_1+f_2)\tilde{W}_{Tx}^{\mathbf{m}*}(f+f_1) g^{\mathbf{i}}(f) \braket{n_3|m_3}\braket{m_3|i_3}df_1  df_2 df\Bigg)+\Bigg(\iiint_{-\infty}^{\infty} \braket{\eta(f,f_1,f_2)|\eta(f,f_1,f_2)} \tilde{W}_{Tx}^{\mathbf{k}*}(f+f_1+f_2) \tilde{W}_{Tx}^{\mathbf{k}}(f+f_2) g^{\mathbf{i}*}(f) \\
&\tilde{W}_{Tx}^{\mathbf{m}}(f+f_1) \braket{k_3|k_3}\braket{i_3|m_3}  \tilde{W}_{Tx}^{\mathbf{m}*}(f+f_2) \tilde{W}_{Tx}^{\mathbf{m}}(f+f_1+f_2) \tilde{W}_{Tx}^{\mathbf{n}*}(f+f_1)  g^{\mathbf{i}}(f) \braket{m_3|m_3}\braket{n_3|i_3}df_1  df_2 df+\iiint_{-\infty}^{\infty} \braket{\eta(f,f_1,f_2)|\eta(f,f_1,f_2)}\\
& \tilde{W}_{Tx}^{\mathbf{k}*}(f+f_1+f_2) \tilde{W}_{Tx}^{\mathbf{k}}(f+f_2) g^{\mathbf{i}*}(f) \tilde{W}_{Tx}^{\mathbf{m}}(f+f_1) \braket{k_3|k_3}\braket{i_3|m_3} \tilde{W}_{Tx}^{\mathbf{n}*}(f+f_2) \tilde{W}_{Tx}^{\mathbf{m}}(f+f_1+f_2) \tilde{W}_{Tx}^{\mathbf{m}*}(f+f_1) g^{\mathbf{i}}(f) \braket{n_3|m_3}\braket{m_3|i_3}df_1  df_2 df\Bigg)
\end{aligned}\end{equation}}
\setcounter{equation}{\value{MYtempeqncnt}}
\hrulefill
\vspace*{4pt}
\end{figure*}
\newcounter{MYtempeqncnt7}
\begin{figure*}[!t]
\normalsize
\setcounter{MYtempeqncnt}{\value{equation}}
\setcounter{equation}{37}

\footnotesize{\begin{equation}\label{eq:38}\begin{aligned}
{}&\sigma_{FON,i_2,i_3}^2=\sum_{{\mathbf{k,m,n}}}\kappa_2^{({\mathbf{k}})} \kappa_1^{({\mathbf{n}})}\Bigg( \iiint_{-\infty}^{\infty}  \braket{\eta(f,f_1,f_2)| \eta(f,f_1,f_2)} \bigg|\tilde{W}_{Tx}^{\mathbf{k}*}(f+f_1+f_2) \tilde{W}_{Tx}^{\mathbf{k}}(f+f_2) g^{\mathbf{i}*}(f) \tilde{W}_{Tx}^{\mathbf{n}}(f+f_1) \braket{k_3|k_3} \braket{i_3|n_3}\\
& +\tilde{W}_{Tx}^{\mathbf{k}*}(f+f_1+f_2) \tilde{W}_{Tx}^{\mathbf{n}}(f+f_2)  g^{\mathbf{i}*}(f) \tilde{W}_{Tx}^{\mathbf{k}}(f+f_1) \braket{k_3|n_3} \braket{i_3|k_3}\bigg|^2 df_1  df_2 df+\iiint_{-\infty}^{\infty} \braket{\eta(f,f_1,f_2)| \eta(f,f_1,f_2)} \bigg|\tilde{W}_{Tx}^{\mathbf{n}*}(f+f_1+f_2)\\
& \tilde{W}_{Tx}^{\mathbf{k}}(f+f_2) g^{\mathbf{i}*}(f) \tilde{W}_{Tx}^{\mathbf{k}}(f+f_1) \braket{n_3|k_3}\braket{i_3|k_3} \bigg|^2 df_1  df_2 df \Bigg)+ \bigg[ \iiint_{-\infty}^{\infty} \braket{\eta(f,f_1,f_2)| \eta(f,f_1,f_2)} \tilde{W}_{Tx}^{\mathbf{k}*}(f+f_1+f_2) \tilde{W}_{Tx}^{\mathbf{k}}(f+f_2) g^{\mathbf{i}*}(f)\\
& \tilde{W}_{Tx}^{\mathbf{k}}(f+f_1) \braket{k_3|k_3} \braket{i_3|k_3} \tilde{W}_{Tx}^{\mathbf{k}*}(f+f_2)  \tilde{W}_{Tx}^{\mathbf{n}}(f+f_1+f_2) \tilde{W}_{Tx}^{\mathbf{n}*}(f+f_1) g^{\mathbf{i}}(f) \braket{n_3|k_3} \braket{n_3|i_3} df_1  df_2 df+\iiint_{-\infty}^{\infty} \braket{\eta(f,f_1,f_2)| \eta(f,f_1,f_2)} \\
& \tilde{W}_{Tx}^{\mathbf{k}*}(f+f_1+f_2) \tilde{W}_{Tx}^{\mathbf{k}}(f+f_2) g^{\mathbf{i}*}(f) \tilde{W}_{Tx}^{\mathbf{k}}(f+f_1) \braket{k_3|k_3} \braket{i_3|k_3} \tilde{W}_{Tx}^{\mathbf{n}*}(f+f_2) \tilde{W}_{Tx}^{\mathbf{n}}(f+f_1+f_2) \tilde{W}_{Tx}^{\mathbf{k}*}(f+f_1) g^{\mathbf{i}}(f) \braket{n_3|n_3} \braket{i_3|k_3} df_1  df_2 df \bigg]\\
&+\bigg[\iiint_{-\infty}^{\infty} \braket{\eta(f,f_1,f_2)| \eta(f,f_1,f_2)} \tilde{W}_{Tx}^{\mathbf{n}*}(f+f_1+f_2) \tilde{W}_{Tx}^{\mathbf{n}}(f+f_2)  g^{\mathbf{i}*}(f) \tilde{W}_{Tx}^{\mathbf{k}}(f+f_1) \braket{n_3|n_3} \braket{i_3|k_3}\tilde{W}_{Tx}^{\mathbf{k}*}(f+f_2) \tilde{W}_{Tx}^{\mathbf{k}}(f+f_1+f_2) \\
&  \tilde{W}_{Tx}^{\mathbf{k}*}(f+f_1) g^{\mathbf{i}}(f) \braket{k_3|k_3} \braket{i_3|k_3} df_1  df_2 df\bigg]+\bigg[\iiint_{-\infty}^{\infty} \braket{\eta(f,f_1,f_2)| \eta(f,f_1,f_2)}  \tilde{W}_{Tx}^{\mathbf{n}*}(f+f_1+f_2) \tilde{W}_{Tx}^{\mathbf{k}}(f+f_2)  g^{\mathbf{i}*}(f) \tilde{W}_{Tx}^{\mathbf{n}}(f+f_1) \braket{k_3|n_3}\\
&  \braket{i_3|n_3} \tilde{W}_{Tx}^{\mathbf{k}*}(f+f_2)\tilde{W}_{Tx}^{\mathbf{k}}(f+f_1+f_2)  \tilde{W}_{Tx}^{\mathbf{k}*}(f+f_1) g^{\mathbf{i}}(f) \braket{k_3|k_3} \braket{i_3|k_3} df_1  df_2 df \bigg]
\end{aligned}\end{equation}}
\setcounter{equation}{\value{MYtempeqncnt}}
\hrulefill
\vspace*{4pt}
\end{figure*}
\newcounter{MYtempeqncnt8}
\begin{figure*}[!t]
\normalsize
\setcounter{MYtempeqncnt}{\value{equation}}
\setcounter{equation}{38}

\small{\begin{equation}\label{eq:39}\begin{aligned}
{}&\sigma_{HON,i_2,i_3}^2=\sum_{{\mathbf{n}}} \kappa_3^{({\mathbf{n}})}
\iiint_{-\infty}^{\infty} \braket{\eta(f,f_1,f_2)| \eta(f,f_1,f_2)} \bigg|\tilde{W}_{Tx}^{\mathbf{n}*}(f+f_1+f_2) \tilde{W}_{Tx}^{\mathbf{n}}(f+f_2)  g^{\mathbf{i}*}(f) \tilde{W}_{Tx}^{\mathbf{n}}(f+f_1) \braket{n_3|n_3}\braket{i_3|n_3}\bigg|^2 df_1  df_2 df
\end{aligned}\end{equation}}
\setcounter{equation}{\value{MYtempeqncnt}}
\hrulefill
\vspace*{4pt}
\end{figure*}

\addtocounter{equation}{12}
\small{\begin{equation}\label{eq:40}\begin{aligned}
{}&\kappa_1=\mu_2\\
&\kappa_2=\mu_4-2\mu_2^2\\
&\kappa_3=\mu_6-4\mu_4\mu_2+12\mu_2^3,\\
&\mu_n=E[|\zeta_{\mathbf{k}}|^n].
\end{aligned}\end{equation}}

\begingroup
\tabcolsep = 8.0pt
\def\arraystretch{0.75}
\begin{table}
 \centering
   \caption{Valid combinations yielding non-zero {\small$E[\zeta_{\mathbf{k}}^*\zeta_{\mathbf{m}}\zeta_{\mathbf{n}}\zeta_{\mathbf{l}}\zeta_{\mathbf{j}}^*\zeta_{\mathbf{o}}^*]$}.}
   \begin{tabular}{| c | c | c | c | c | c | c |}
    \hline
     {\footnotesize $\mathbf{k}^*$} &{\footnotesize $\mathbf{m}$} & {\footnotesize $\mathbf{n}$}&{\footnotesize $\mathbf{l}$} & {\footnotesize $\mathbf{j}^*$} & {\footnotesize $\mathbf{o}^*$}& \\ \hline
     {\footnotesize $*$} &{\footnotesize $*$} & {\footnotesize $*$} &{\footnotesize $*$} &{\footnotesize $*$} &{\footnotesize $*$} &{\footnotesize $HON$}\\ \hline
     {\footnotesize $*$} &{\footnotesize $*$} & {\footnotesize $**$} &{\footnotesize $*$} &{\footnotesize $*$} &{\footnotesize $**$} &{\footnotesize $FONa$}\\ \hline
     {\footnotesize $*$} &{\footnotesize $*$} & {\footnotesize $**$} &{\footnotesize $*$} &{\footnotesize $**$} &{\footnotesize $*$} &{\footnotesize $FONa$}\\ \hline
     {\footnotesize $*$} &{\footnotesize $**$} & {\footnotesize $*$} &{\footnotesize $*$} &{\footnotesize $*$} &{\footnotesize $**$} &{\footnotesize $FONa$}\\ \hline
     {\footnotesize $*$} &{\footnotesize $**$} & {\footnotesize $*$} &{\footnotesize $*$} &{\footnotesize $**$} &{\footnotesize $*$} &{\footnotesize $FONa$}\\ \hline
     {\footnotesize $*$} &{\footnotesize $*$} & {\footnotesize $*$} &{\footnotesize $**$} &{\footnotesize $*$} &{\footnotesize $**$} &{\footnotesize $FONb$}\\ \hline
     {\footnotesize $*$} &{\footnotesize $*$} & {\footnotesize $*$} &{\footnotesize $**$} &{\footnotesize $**$} &{\footnotesize $*$} &{\footnotesize $FONb$}\\ \hline
     {\footnotesize $**$} &{\footnotesize $*$} & {\footnotesize $**$} &{\footnotesize $*$} &{\footnotesize $*$} &{\footnotesize $*$} &{\footnotesize $FONc$}\\ \hline
     {\footnotesize $**$} &{\footnotesize $*$} & {\footnotesize $*$} &{\footnotesize $**$} &{\footnotesize $*$} &{\footnotesize $*$} &{\footnotesize $FONc$}\\ \hline
     {\footnotesize $**$} &{\footnotesize $*$} & {\footnotesize $*$} &{\footnotesize $**$} &{\footnotesize $*$} &{\footnotesize $*$} &{\footnotesize $FONd$}\\ \hline
     {\footnotesize $*$} &{\footnotesize $*$} & {\footnotesize $**$} &{\footnotesize $+$} &{\footnotesize $+$} &{\footnotesize $**$} &{\footnotesize $GNa$}\\ \hline
     {\footnotesize $*$} &{\footnotesize $*$} & {\footnotesize $**$} &{\footnotesize $+$} &{\footnotesize $**$} &{\footnotesize $+$} &{\footnotesize $GNa$}\\ \hline
     {\footnotesize $*$} &{\footnotesize $**$} & {\footnotesize $*$} &{\footnotesize $+$} &{\footnotesize $+$} &{\footnotesize $**$} &{\footnotesize $GNa$}\\ \hline
     {\footnotesize $*$} &{\footnotesize $**$} & {\footnotesize $*$} &{\footnotesize $+$} &{\footnotesize $**$} &{\footnotesize $+$} &{\footnotesize $GNa$}\\ \hline
     {\footnotesize $*$} &{\footnotesize $+$} & {\footnotesize $**$} &{\footnotesize $*$} &{\footnotesize $+$} &{\footnotesize $**$} &{\footnotesize $GNb$}\\ \hline
     {\footnotesize $*$} &{\footnotesize $+$} & {\footnotesize $**$} &{\footnotesize $*$} &{\footnotesize $**$} &{\footnotesize $+$} &{\footnotesize $GNb$}\\ \hline
\end{tabular}
\label{table:1}
\end{table}
\endgroup

Table \ref{table:1} shows valid combinations yielding non-zero {\small$E[\zeta_{\mathbf{k}}^*\zeta_{\mathbf{m}}\zeta_{\mathbf{n}}\zeta_{\mathbf{l}}\zeta_{\mathbf{j}}^*\zeta_{\mathbf{o}}^*]$} where FONb, FONc, and GNa are the removed terms from the EGN model formulation due to the CPE assumption.
Moreover, {\small$\sigma_{EGN}^{(i_3,odd)}(f) = \sigma_{EGN}^{(i_3,even)}(f)$}, and {\small$\sigma_{EGN}^{(p)}(f)=\sigma_{EGN}^{(i_3,odd)}(f)+\sigma_{EGN}^{(i_3,even)}(f)$}. Therefore, the GN, FON, and HON contributions of the nonlinear noise variance of {\small$i_2$}th channel and {\small $p$}th mode can be written as (\ref{eq:41}), (\ref{eq:42}), and (\ref{eq:43}), respectively.

\newcounter{MYtempeqncnt9}
\begin{figure*}[!t]
\normalsize
\setcounter{MYtempeqncnt}{\value{equation}}
\setcounter{equation}{40}

\small{\begin{equation}\label{eq:41}\begin{aligned}
{}&\sigma_{GN,i_2,p}^2=\sum_{q=1}^{D} 3\sum_{k_2,m_2,n_2} \kappa_1^{(k_2)}\kappa_1^{(m_2)}\kappa_1^{(n_2)}
\iiint_{-\infty}^{\infty} |\eta(f,f_1,f_2)|^2 \bigg(
\tilde{G}_{Tx}^{m_2,q}(f+f_2) \tilde{G}_{Tx}^{k_2,q}(f+f_1+f_2) \tilde{G}_{Tx}^{n_2,p}(f+f_1) g^{i_2,p}(f)+\\
&\tilde{G}_{Tx}^{n_2,q}(f+f_2) \tilde{G}_{Tx}^{k_2,q}(f+f_1+f_2) \tilde{G}_{Tx}^{ m_2,p}(f+f_1) g^{i_2,p}(f)\bigg) df_1  df_2 df
\end{aligned}\end{equation}}
\setcounter{equation}{\value{MYtempeqncnt}}
\hrulefill
\vspace*{4pt}
\end{figure*}
\newcounter{MYtempeqncnt10}
\begin{figure*}[!t]
\normalsize
\setcounter{MYtempeqncnt}{\value{equation}}
\setcounter{equation}{41}

\small{\begin{equation}\label{eq:42}\begin{aligned}
{}&\sigma_{FON,i_2,p}^2=\sum_{q=1}^{D} \sum_{k_2,n_2}\kappa_2^{(k_2)} \kappa_1^{(n_2)}
\Bigg(\iiint_{-\infty}^{\infty} |\eta(f,f_1,f_2)|^2 5 \tilde{G}_{Tx}^{ k_2,q}(f+f_1+f_2) \tilde{G}_{Tx}^{k_2,q}(f+f_2) g^{i_2,p}(f) \tilde{G}_{Tx}^{n_2,p}(f+f_1) df_1  df_2 df +\\
& \iiint_{-\infty}^{\infty} |\eta(f,f_1,f_2)|^2 \tilde{G}_{Tx}^{n_2,q}(f+f_1+f_2) \tilde{G}_{Tx}^{k_2,q}(f+f_2) g^{i_2,p}(f) \tilde{G}_{Tx}^{k_2,p}(f+f_1) df_1  df_2 df \Bigg)
\end{aligned}\end{equation}}
\setcounter{equation}{\value{MYtempeqncnt}}
\hrulefill
\vspace*{4pt}
\end{figure*}
\newcounter{MYtempeqncnt11}
\begin{figure*}[!t]
\normalsize
\setcounter{MYtempeqncnt}{\value{equation}}
\setcounter{equation}{42}

\small{\begin{equation}\label{eq:43}\begin{aligned}
{}&\sigma_{HON,i_2,p}^2=\sum_{q=1}^{D} \sum_{n_2}\kappa_3^{(n_2)}
\iiint_{-\infty}^{\infty} |\eta(f,f_1,f_2)|^2 \tilde{G}_{Tx}^{n_2,q}(f+f_1+f_2) \tilde{G}_{Tx}^{n_2,q}(f+f_2) g^{i_2,p}(f) \tilde{G}_{Tx}^{n_2,p}(f+f_1) df_1  df_2 df
\end{aligned}\end{equation}}
\setcounter{equation}{\value{MYtempeqncnt}}
\hrulefill
\vspace*{4pt}
\end{figure*}
The power spectral density of the optical launched signal can be written as  {\small$\tilde{G}_{Tx}^{i_2,p}(f)=P_{i_2,p}g_{i_2,p}(f)$}. Accordingly, the GN, FON, and HON contributions of the nonlinear noise variance of {\small$i_2$}th channel and {\small$p$}th mode can be expressed as (\ref{eq:44}), (\ref{eq:45}), and (\ref{eq:46}), respectively.
\newcounter{MYtempeqncnt12}
\begin{figure*}[!t]
\normalsize
\setcounter{MYtempeqncnt}{\value{equation}}
\setcounter{equation}{43}

\small{\begin{equation}\label{eq:44}\begin{aligned}
{}&\sigma_{GN,i_2,p}^2=\sum_{q=1}^{D} 3/4\sum_{k_2,m_2,n_2}\kappa_1^{(k_2)}\kappa_1^{(m_2)}\kappa_1^{(n_2)} \iiint_{-\infty}^{\infty} |\eta(f,f_1,f_2)|^2 \bigg(P_{k_2,q} P_{ m_2,q} P_{n_2,p} g^{m_2,q}(f+f_2) g^{k_2,q}(f+f_1+f_2)\\
&g^{n_2,p}(f+f_1) g^{i_2,p}(f)+ P_{k_2,q} P_{ m_2,p} P_{n_2,q} g^{n_2,q}(f+f_2) g^{k_2,q}(f+f_1+f_2) g^{ m_2,p}(f+f_1) g^{i_2,p}(f)\bigg) df_1  df_2 df
\end{aligned}\end{equation}}
\setcounter{equation}{\value{MYtempeqncnt}}
\hrulefill
\vspace*{4pt}
\end{figure*}
\newcounter{MYtempeqncnt13}
\begin{figure*}[!t]
\normalsize
\setcounter{MYtempeqncnt}{\value{equation}}
\setcounter{equation}{44}

\small{\begin{equation}\label{eq:45}\begin{aligned}
{}&\sigma_{FON,i_2,p}^2=\sum_{q=1}^{D} 1/4\sum_{k_2,n_2} \kappa_2^{(k_2)} \kappa_1^{(n_2)} \Bigg(P_{ k_2,q}^2 P_{n_2,p}
5 \iiint_{-\infty}^{\infty} |\eta(f,f_1,f_2)|^2 g^{ k_2,q}(f+f_1+f_2) g^{k_2,q}(f+f_2) g^{i_2,p}(f) g^{n_2,p}(f+f_1) \\
& df_1  df_2 df + P_{ k_2,p}P_{ k_2,q} P_{n_2,q} \iiint_{-\infty}^{\infty} |\eta(f,f_1,f_2)|^2 g^{n_2,q}(f+f_1+f_2) g^{k_2,q}(f+f_2) g^{i_2,p}(f) g^{k_2,p}(f+f_1) df_1  df_2 df \Bigg)
\end{aligned}\end{equation}}
\setcounter{equation}{\value{MYtempeqncnt}}
\hrulefill
\vspace*{4pt}
\end{figure*}
\newcounter{MYtempeqncnt14}
\begin{figure*}[!t]
\normalsize
\setcounter{MYtempeqncnt}{\value{equation}}
\setcounter{equation}{45}

\small{\begin{equation}\label{eq:46}\begin{aligned}
\sigma_{HON,i_2,p}^2=\sum_{q=1}^{D} 1/4\sum_{n_2}\kappa_3^{(n_2)}  P_{n_2,q}^2 P_{n_2,p}
\iiint_{-\infty}^{\infty} |\eta(f,f_1,f_2)|^2 g^{n_2,q}(f+f_1+f_2) g^{n_2,q}(f+f_2) g^{i_2,p}(f) g^{n_2,p}(f+f_1) df_1  df_2 df
\end{aligned}\end{equation}}
\setcounter{equation}{\value{MYtempeqncnt}}
\hrulefill
\vspace*{4pt}
\end{figure*}
Therefore, the nonlinear noise variance can be written as (\ref{eq:47}).
\newcounter{MYtempeqncnt15}
\begin{figure*}[!t]
\normalsize
\setcounter{MYtempeqncnt}{\value{equation}}
\setcounter{equation}{46}

\small{\begin{equation}\label{eq:47}\begin{aligned}
{}&\sigma_{EGN,i_2,p}^2=\sum_{q=1}^{D} 3/4\sum_{k_2,m_2,n_2}\kappa_1^{(k_2)}\kappa_1^{(m_2)}\kappa_1^{(n_2)} P_{k_2,q} P_{m_2,q} P_{n_2,p}X_{i_2,p}^{a}(k_2,m_2,n_2,q)+1/4\sum_{k_2,n_2}\kappa_2^{(k_2)} \kappa_1^{(n_2)} (P_{ k_2,q}^2 P_{n_2,p} 5X_{i_2,p}^{b}(k_2,k_2,n_2,q)\\
&+ P_{k_2,p} P_{k_2,q}P_{n_2,q}X_{i_2,p}^{c}(k_2,n_2,k_2,q))+1/4\sum_{n_2}\kappa_3^{(n_2)}  P_{n_2,q}^2P_{n_2,p} X_{i_2,p}^{d}(n_2,n_2,n_2,q)\\
&where \left\{
    \begin{array}{ll}
X_{i_2,p}^{a}(k_2,m_2,n_2,q)=\iiint_{-\infty}^{\infty} |\eta(f,f_1,f_2)|^2g^{m_2,q}(f+f_2) g^{k_2,q}(f+f_1+f_2) g^{n_2,p}(f+f_1) g^{i_2,p}(f) df_1 df_2 df\\
X_{i_2,p}^{b}(k_2,k_2,n_2,q)= \iiint_{-\infty}^{\infty} |\eta(f,f_1,f_2)|^2 g^{ k_2,q}(f+f_1+f_2) g^{k_2,q}(f+f_2) g^{i_2,p}(f) g^{n_2,p}(f+f_1) df_1  df_2 df \\
X_{i_2,p}^{c}(k_2,n_2,k_2,q)= \iiint_{-\infty}^{\infty} |\eta(f,f_1,f_2)|^2 g^{n_2,q}(f+f_1+f_2) g^{k_2,q}(f+f_2) g^{i_2,p}(f) g^{k_2,p}(f+f_1) df_1  df_2 df\\
X_{i_2,p}^{d}(n_2,n_2,n_2,q)=\iiint_{-\infty}^{\infty} |\eta(f,f_1,f_2)|^2 g^{n_2,q}(f+f_1+f_2) g^{n_2,q}(f+f_2) g^{i_2,p}(f) g^{n_2,p}(f+f_1) df_1  df_2 df
    \end{array}
\right.
\end{aligned}\end{equation}}
\setcounter{equation}{\value{MYtempeqncnt}}
\hrulefill
\vspace*{4pt}
\end{figure*}

\section*{Appendix B: Convexity proof of (33)}
The expression (\ref{eq:48})
\newcounter{MYtempeqncnt29}
\begin{figure*}[!t]
\normalsize
\setcounter{MYtempeqncnt}{\value{equation}}
\setcounter{equation}{47}

\small{\begin{equation}\label{eq:48}\begin{aligned}
{}&log(\prod_{n=1}^{N_s}(e^{g_n}L_n) (F (G_{BA}-1) h \nu B_{i_2})+\sum_{s=1}^{N_s}[(F(e^{g_s}-1)h\nu B_{i_2})\prod_{n=s+1}^{N_s}(e^{g_n}L_n)]
 + \sum_{l_1,l_2,l_3=1}^{D N_{ch}} \kappa_1^{{'}^{(l_1)}}\kappa_1^{{'}^{(l_2)}}\kappa_1^{{'}^{(l_3)}} e^{\hat{P}_{l_1}+\hat{P}_{l_2}+\hat{P}_{l_3}}3H_{l}^{a}(l_1,l_2,l_3)
\\
 &+ \sum_{l_1,l_2,l_3=1}^{D N_{ch}}  \kappa_2^{{'}^{(l_1)}}\kappa_1^{{'}^{(l_2)}}(e^{2\hat{P}_{l_1}+\hat{P}_{l_2}}5H_{l}^{b}(l_1,l_1,l_2)+e^{\hat{P}_{l_1}+\hat{P}_{l_2}+\hat{P}_{l_3}}H_{l}^{c}(l_1,l_2,l_3))
+\sum_{l_1,l_2=1}^{D N_{ch}} \kappa_3^{{'}^{(l_1)}} e^{2\hat{P}_{l_1}+\hat{P}_{l_2}} H_{l}^{d}(l_1,l_1,l_2)
\bigg)-(\hat{P}_{l}+\sum_{n=1}^{N_s}g_n log(L_n))
 \end{aligned}\end{equation}}
 \setcounter{equation}{\value{MYtempeqncnt}}
\hrulefill
\vspace*{4pt}
\end{figure*}
is convex in {\small$\hat{P}_{l}, g_s$}, since {\small$log-sum-exp (x)$} is a convex function in {\small$x$}. The constraint function of (\ref{eq:33}) the summation of some convex functions, therefore, it is convex. The objective and constraint functions of (\ref{eq:33}) are convex, therefore, (\ref{eq:33}) is a convex optimization problem.

\smallskip

\begin{backmatter}
\bmsection{\Large{Disclosures}} The authors declare no conflicts of interest.
\end{backmatter}

\end{document}